\documentclass[journal=jctcce,manuscript=article]{achemso}

\usepackage[utf8]{inputenc} % set input encoding (not needed with XeLaTeX)

\usepackage{geometry} % to change the page dimensions
\usepackage{graphicx} % support the \includegraphics command and options

\usepackage{booktabs} % for much better looking tables
\usepackage{array} % for better arrays (eg matrices) in maths
\usepackage{paralist} % very flexible & customisable lists (eg. enumerate/itemize, etc.)
\usepackage{verbatim} % adds environment for commenting out blocks of text & for better verbatim
\usepackage{subfig} % make it possible to include more than one captioned figure/table in a single float
\usepackage{amsmath}
\usepackage{amssymb}
\usepackage{amsthm}
\usepackage{amsfonts}
\usepackage{listings}
\usepackage{color}
\usepackage{amsfonts}
\usepackage{caption}
\usepackage{listings}
\usepackage{amscd}
\usepackage{tikz}
\usetikzlibrary {shapes.geometric} 

\lstdefinelanguage{GAP}{%
  morekeywords={%
    Assert,Info,IsBound,QUIT,%
    TryNextMethod,Unbind,and,break,%
    continue,do,elif,%
    else,end,false,fi,for,%
    function,if,in,local,%
    mod,not,od,or,%
    quit,rec,repeat,return,%
    then,true,until,while%
  },%
  sensitive,%
  morecomment=[l]\#,%
  morestring=[b]",%
  morestring=[b]',%
}[keywords,comments,strings]

\lstdefinelanguage{bash}{%
  morekeywords={%
    Assert,Info,IsBound,QUIT,%
    TryNextMethod,Unbind,and,break,%
    continue,do,elif,%
    else,end,false,fi,for,%
    function,if,in,local,%
    mod,not,od,or,%
    quit,rec,repeat,return,%
    then,true,until,while%
  },%
  sensitive,%
  morecomment=[l]\#,%
  morestring=[b]",%
  morestring=[b]',%
}[keywords,comments,strings]

\usepackage[T1]{fontenc}
\usepackage[variablett]{lmodern}
\usepackage{xcolor}
\newtheorem{theorem}{Theorem}[section]
\newtheorem{lemma}{Lemma}[section]
\newtheorem{definition}{Definition}[section]

\usepackage{fancyhdr} % This should be set AFTER setting up the page geometry
\usepackage[nottoc,notlof,notlot]{tocbibind} % Put the bibliography in the ToC
\usepackage[titles,subfigure]{tocloft} % Alter the style of the Table of Contents

\usetikzlibrary{shapes.geometric, arrows}

\tikzstyle{startstop} = [rectangle, rounded corners, 
minimum width=3cm, 
minimum height=1cm,
text centered, 
draw=black, 
fill=red!30]

\tikzstyle{io} = [trapezium, 
trapezium stretches=true, % A later addition
trapezium left angle=70, 
trapezium right angle=110, 
minimum width=3cm, 
minimum height=1cm, text centered, 
draw=black, fill=blue!30]

\tikzstyle{process} = [rectangle, 
minimum width=3cm, 
minimum height=1cm, 
text width=15cm, 
text centered,  
draw=black, 
fill=orange!30]

\tikzstyle{decision} = [diamond, 
minimum width=3cm, 
minimum height=1cm, 
text centered, 
draw=black, 
fill=green!30]
\tikzstyle{arrow} = [thick,->,>=stealth]

\title{Reproducing Reaction Route Map on the Shape Space from its Quotient by Complete Nuclear Permutation-Inversion group}
\author{Hiroshi Teramoto}
\affiliation[Kansai Univ]
{Faculty of Engineering Science, Kansai University, Suita 564-8680, Japan}
\email{teramoto@kansai-u.ac.jp}

\author{Takuya Saito}
\affiliation[Math Sci Hokkaido Univ]
{Department of Mathematics, Faculty of Science, Hokkaido University, Sapporo 060-0810, Japan}
\alsoaffiliation[Unito]
{Department of Economics and Statistics, University of Turin, 10124 Turin, Italy}

\author{Masamitsu Aoki}
\affiliation[Math Sci Hokkaido Univ]
{Department of Mathematics, Faculty of Science, Hokkaido University, Sapporo 060-0810, Japan}

\author{Burai Murayama}
\affiliation[Grad Chem Sci Eng Hokkaido Univ]
{Graduate School of Chemical Sciences and Engineering, Hokkaido University, Sapporo 060-0810, Japan}

\author{Masato Kobayashi}
\affiliation[Chem Sci Hokkaido Univ]
{Department of Chemistry, Faculty of Science, Hokkaido University, Sapporo 060-0810, Japan}
\alsoaffiliation[ICReDD Hokkaido Univ]
{WPI-ICReDD, Hokkaido University, Sapporo 001-0021, Japan}

\author{Takenobu Nakamura}
\affiliation[AIST]
{National Institute of Advanced Industrial Science and Technology, Tsukuba 305-8568, Japan}

\author{Tetsuya Taketsugu}
\affiliation[Chem Sci Hokkaido Univ]
{Department of Chemistry, Faculty of Science, Hokkaido University, Sapporo 060-0810, Japan}
\alsoaffiliation[ICReDD Hokkaido Univ]
{WPI-ICReDD, Hokkaido University, Sapporo 001-0021, Japan}

\date{} % Activate to display a given date or no date (if empty),
\begin{document}
\maketitle
\begin{abstract}
This study develops an algorithm to reproduce reaction route maps (RRMs) in shape space from the outputs of potential search algorithms. To demonstrate the algorithm, GRRM is utilized as a potential search algorithm but the proposed algorithm should work with other potential search algorithms in principle. The proposed algorithm does not require any encoding of the molecular configurations and is thus applicable to complicated realistic molecules for which efficient encoding is not readily available. We show subgraphs of an RRM mapped to each other by the action of the symmetry group are isomorphic and also provide an algorithm to compute the set of feasible transformations in the sense of Longuet--Higgins. We demonstrate the proposed algorithm in toy models and in more realistic molecules. Finally, we remark on absolute rate theory from our perspective.
\end{abstract}

\section{Introduction}
In chemistry, the potential energy functions of molecules play an important role in understanding their statistical and dynamical properties of the molecules. The potential energy function of a molecule is defined on the shape space of the molecule \cite{PhysRevA.52.2035}, i.e., the configuration space of the atoms composing the molecule in the three-dimensional (3D) space in which two configurations are identified if one of the two configurations can be matched with the other by 3D spatial translation and rotation.

Important characteristics of the potential energy function are its equilibrium and transition states, and the connection among them through reaction paths. There are extensive studies on algorithms to search such equilibrium and transition states, and the reaction paths connecting them. To make searching algorithms efficient, it is important to avoid rediscovering known equilibrium and stransition states as reviewed in Chapter~8 in Ref.~\citenum{PetersBook}. Therefore, typically in these algorithms, two conformations of a molecule in the shape space are identified if one of them can be matched to the other by the spatial inversion and permutations of identical atoms, i.e., by the action of the complete nuclear permutation-inversion (CNPI) group. The resulting reaction route map (RRM) is the quotient of the RRM in the shape space by CNPI group. For instance, the global reaction route mapping (GRRM) program \cite{C3CP44063J, https://doi.org/10.1002/jcc.25106} is one such a program \cite{Ohno_Book2022}.

Obtaining the RRM of a molecule in the shape space proves useful for at least the following three reasons: First, it enables the computation of the set of feasible transformations of the molecule, that is, the subset of CNPI transformations that can be achieved without overcoming an insurmountable energy barrier\cite{Longuet-Higgins1963}. The RRM in the shape space is mandatory to compute the set of feasible transformations of the molecule, as it requires knowledge of which isomers are mutually energetically accessible for a given energy. Several studies have been conducted on feasible transformations and their application to the tunneling splitting of the spectra of permutational isomers\cite{Brocas1983,doi:10.1063/1.1730820}. Second, RRMs are also used to understand how dynamics proceed \cite{D1CC04667E,Tsutsumi2018}. For that purpose, RRMs in shape space offer a more intuitive interpretation of dynamics than ones in symmetry-reduced space, a point highlighted by Mezey\cite{MezeyBook}. Third, RRMs in shape space provide a fair basis for comparing descriptors across molecules. Recently, the persistent homology\cite{Mirth2021_EL-PH,Murayama2022} and disconnectivity graph\cite{Becker1997_Disconnectivity-graph,Wales2005_Disconnectivity-graph} of an RRM (whether in symmetry-reduced space 
\begin{math}
\left( \mathbb{R}^3 \right)^N / \mathrm{E}^+ \left( 3 \right) \times \mathrm{Sym}_\Omega
\end{math}
or shape space) have been used as descriptors of a molecule to characterize its chemical properties. If one wants to compare such descriptors across different molecules, an RRM in shape space should be used since different molecules have different symmetries, and using an RRM in symmetry-reduced space might result in an unfair comparison.

In this study, an algorithm to reproduce RRMs in shape space from the outputs of potential search algorithms was developed. To demonstrate the proposed algorithm, we use GRRM as a potential search algorithm but the proposed algorithm should work with other potential search algorithms in principle. The remainder of this paper is organized as follows. Section~2 introduces the terminology and setting used herein, along with a summary of the previous results. Section~3 presents an algorithm based on these settings. Section~4 demonstrates that subgraphs of an RRM mapped to each other by the action of the symmetry group are isomorphic and provide an algorithm to compute the set of feasible permutations. Section~5 and 6 demonstrate the proposed algorithm using toy models and for more realistic molecules, respectively. Section~7 discusses the absolute rate theory from our perspective. Finally, Section~8 concludes the paper and provides future perspectives.
\section{Settings of Potential Energy Surface}
Let \(\Omega\) be the set of atoms in the system, where the system comprises \(N\) atoms of \(l\) types chemical elements. Let \(r_i\) be the mass-weighted coordinate of the $i$-th atom in \(\mathbb{R}^3\), and \(r\) denotes their \(N\)-tuple \(\left( r_1, \cdots, r_N \right) \in \left( \mathbb{R}^3 \right)^N\).
Let \(\mathrm{Sym}_\Omega\) be the group consisting of all the permutations of the atoms of the same types.
Note that this group is isomorphic to the direct product group of the symmetric groups.
For instance, suppose that
\begin{math}
n_j
\end{math}
denotes the number of atoms of the $j$-th type for
\begin{math}
j \in \left\{ 1, \cdots, l \right\}
\end{math}. As the total number of atoms is
\begin{math}
N
\end{math},
\begin{math}
N = n_1 + \cdots + n_l
\end{math}
holds. In this case,
\begin{math}
\mathrm{Sym}_\Omega \cong \mathfrak{S}_{n_1} \times \cdots \times \mathfrak{S}_{n_l}
\end{math}
holds true, where \(\mathfrak{S}_n\) denotes the symmetric group of degree \(n\). For details on CNPI and symmetric groups, see Ref.~\citenum{Bunker_Book1979}.
Let
\begin{math}
\mathrm{O} \left( 3 \right)
\end{math}
be the group of
\begin{math}
3 \times 3
\end{math}
orthogonal matrices,
\begin{math}
\mathrm{SO} \left( 3 \right)
\end{math}
be the subgroup of
\begin{math}
\mathrm{O} \left( 3 \right)
\end{math}
comprising the matrices of the determinant $1$ and
\begin{math}
\mathrm{T} \left( 3 \right)
\end{math}
be 3D translational group and
\begin{math}
\mathrm{E} \left( 3 \right) = \mathrm{T} \left( 3 \right) \rtimes \mathrm{O} \left( 3 \right)
\end{math}
(\begin{math}
\mathrm{E}^+ \left( 3 \right) = \mathrm{T} \left( 3 \right) \rtimes \mathrm{SO} \left( 3 \right)
\end{math})
be their semi-direct product known as the Euclidean group (special Euclidean group). Next, consider the actions of 
\begin{math}
\mathrm{E} \left( 3 \right)
\end{math}
and
\begin{math}
\mathrm{Sym}_\Omega
\end{math}
on 
\begin{math}
\left( \mathbb{R}^3 \right)^N
\end{math}. In the case of 
\begin{math}
\mathrm{E} \left( 3 \right)
\end{math}, 
\begin{equation}
g \cdot r = \left( A r_1 + m_1^{1/2} t, A r_2 + m_2^{1/2} t, \cdots, A r_N + m_N^{1/2} t \right)
\end{equation}
where 
\begin{math}
m_i
\end{math}
is the mass of 
\begin{math}
i
\end{math}-th atom that depends only on the type of the atom,
\begin{math}
g = \left( t, A \right) \in \mathrm{E} \left( 3 \right)
\end{math}
and 
\begin{math}
A r_j
\end{math}
is the matrix product of 
\begin{math}
A
\end{math}
and 
\begin{math}
r_j
\end{math}
for 
\begin{math}
j \in \left\{ 1, \cdots, N \right\}
\end{math}. In the case of 
\begin{math}
\mathrm{Sym}_\Omega
\end{math}, 
\begin{equation}
\sigma \cdot r = \left( r_{\sigma \left( 1 \right)}, r_{\sigma \left( 2 \right)}, \cdots, r_{\sigma \left( N \right)} \right)
\end{equation}
where 
\begin{math}
\sigma \in \textnormal{Sym}_\Omega
\end{math}
and 
\begin{math}
\sigma \left( j \right)
\end{math}
is the image of 
\begin{math}
j
\end{math}
by the permutation 
\begin{math}
\sigma
\end{math}
for 
\begin{math}
j \in \left\{ 1, \cdots, N \right\}
\end{math}. Note that the two actions commute, i.e., 
\begin{math}
\sigma \cdot \left( g \cdot r \right) = g \cdot \left( \sigma \cdot r \right)
\end{math}
for all 
\begin{math}
\sigma \in \mathrm{Sym}_\Omega
\end{math}, 
\begin{math}
g \in \mathrm{E} \left( 3 \right)
\end{math}
and 
\begin{math}
r \in \left( \mathbb{R}^3 \right)^N
\end{math}. Therefore, we can consider the action of the direct product group 
\begin{math}
\mathrm{E} \left( 3 \right) \times \mathrm{Sym}_\Omega
\end{math}
on 
\begin{math}
\left( \mathbb{R}^3 \right)^N
\end{math}
in the obvious way. Let 
\begin{math}
\mathrm{C_i}
\end{math}
be the subgroup of 
\begin{math}
\mathrm{O} \left( 3 \right)
\end{math}
generated by the matrix 
\begin{math}
\mathrm{i} = \left(
\begin{matrix}
-1 & 0 & 0 \\
0 & -1 & 0 \\
0 & 0 & -1
\end{matrix}
\right)
\end{math}. The direct product group 
\begin{math}
\mathrm{E} \left( 3 \right) \times \mathrm{Sym}_\Omega
\end{math}
contains the complete nuclear permutation inversion (CNPI) group
\begin{math}
\mathrm{C_i} \times \mathrm{Sym}_\Omega
\end{math}
as a subgroup, which is introduced by Longuet--Higgins as a symmetric group of non-rigid molecules \cite{Longuet-Higgins1963}.

Now, consider the potential energy function 
\begin{math}
V \colon \left( \mathbb{R}^3 \right)^N \rightarrow \mathbb{R}
\end{math},
which is fourth continuously differentiable, i.e., 
\begin{math}
C^4
\end{math}, 
and is invariant under the action of 
\begin{math}
\mathrm{E} \left( 3 \right)
\end{math}
and 
\begin{math}
\mathrm{Sym}_\Omega
\end{math}, i.e., 
\begin{math}
V \left( g \cdot r \right) = V \left( r \right)
\end{math}
and 
\begin{math}
V \left( \sigma \cdot r \right) = V \left( r \right), 
\end{math}
for all 
\begin{math}
g \in \mathrm{E} \left( 3 \right)
\end{math}, 
\begin{math}
\sigma \in \mathrm{Sym}_\Omega
\end{math}, and 
\begin{math}
r \in \left( \mathbb{R}^3 \right)^N
\end{math}. The assumption that 
\begin{math}
V
\end{math}
is 
\begin{math}
C^4
\end{math}
is necessary to guarantee the unique existence of the unstable manifold of the first--rank saddle of the gradient flow, using Kelley's theorem \cite{Kelley1967}. Note that this assumption is necessary only in neighborhoods of transition states and a weaker regularity condition such as the Lipschitz continuity of the gradient of 
\begin{math}
V
\end{math}
is sufficient along the reaction coordinates outside the neighborhoods. In case of a Born--Oppenheimer potential energy function of a nonrelativistic Sch\"odinger equation, the potential energy function for a nondegenerate electronic state of a molecule is an analytic function of the nuclear coordinates everywhere except points at which the nuclei coincide \cite{AIHPA_1986__45_4_339_0,Ganelin1991}.

Let
\begin{math}
\nabla V \left( r \right) = \left( \frac{\partial V \left( r \right)}{\partial r_1}, \cdots, \frac{\partial V \left( r \right)}{\partial r_N} \right)
\end{math}
and 
\begin{math}
H \left( r \right)
\end{math}
be the gradient and the Hesse matrix of
\begin{math}
V
\end{math}
at
\begin{math}
r = \left( r_1, \cdots, r_N \right)
\end{math}, respectively.
As 
\begin{math}
V
\end{math}
is invariant under the action of 
\begin{math}
\mathrm{O} \left( 3 \right) \subset \mathrm{E} \left( 3 \right)
\end{math}, 
\begin{math}
\nabla V
\end{math}
is 
\begin{math}
\mathrm{O} \left( 3 \right)
\end{math}-equivariant, i.e., 
\begin{math}
g \cdot \nabla V \left( r \right) = \nabla V \left( g \cdot r \right)
\end{math}
for all 
\begin{math}
g \in \mathrm{O} \left( 3 \right)
\end{math}
and 
\begin{math}
r \in \left( \mathbb{R}^3 \right)^N
\end{math}
(in p.~84, Theorem~15 in Ref.~\citenum{Heidrich1991}). Similarly, 
\begin{math}
\sigma \cdot \nabla V \left( r \right) = \nabla V \left( \sigma \cdot r \right)
\end{math}
holds for all 
\begin{math}
\sigma \in \mathrm{Sym}_\Omega
\end{math}
and 
\begin{math}
r \in \left( \mathbb{R}^3 \right)^N
\end{math}. Moreover, for any stationary point 
\begin{math}
r \in \left( \mathbb{R}^3 \right)^N
\end{math}
of 
\begin{math}
V
\end{math}
and 
\begin{math}
    v \in \left( \mathbb{R}^3 \right)^N
\end{math},  
\begin{math}
g \cdot \left( H \left( r \right) v \right) = H \left( g \cdot r \right) \left( g \cdot v \right)
\end{math}
holds true for all 
\begin{math}
g \in \mathrm{O} \left( 3 \right)
\end{math}. Therefore, if 
\begin{math}
e \in \left( \mathbb{R}^3 \right)^N
\end{math}
is an eigenvector of 
\begin{math}
H \left( r \right)
\end{math}, 
\begin{math}
g \cdot e
\end{math}
is an eigenvector of 
\begin{math}
H \left( g \cdot r \right)
\end{math}
of the same eigenvalue for all 
\begin{math}
g \in \mathrm{O} \left( 3 \right)
\end{math} (p.84, Theorem 15 in Ref.~\citenum{Heidrich1991}). 

In this setting, the Hesse matrix of 
\begin{math}
V
\end{math}
at any stationary point 
\begin{math}
r \in \left( \mathbb{R}^3 \right)^N
\end{math}, 
\begin{math}
H \left( r \right)
\end{math}
has 
\begin{itemize}
\item 
\begin{math}
6
\end{math}
zero eigenvalues if the set of points
\begin{math}
r_1, \cdots, r_N
\end{math}
is non-collinear,
\item 
\begin{math}
5
\end{math}
zero eigenvalues if the set of points is collinear and not all the points are in the same position, and
\item 
\begin{math}
3
\end{math}
zero eigenvalues if 
\begin{math}
r_1 = \cdots = r_N
\end{math}.
\end{itemize} In the following, we define equilibrium (transition) state as a stationary point 
\begin{math}
r
\end{math}
of 
\begin{math}
V
\end{math}
at which the Hessian
\begin{math}
H \left( r \right)
\end{math}
does not have zero eigenvalues in addition to the aforementioned zero eigenvalues and all the other nonzero eigenvalues of 
\begin{math}
H \left( r \right)
\end{math}
are positive (positive aside from one).

\section{Constructing  RRM on Shape Space}
In this section, we construct the RRM of a molecule in the shape space 
\begin{math}
\left( \mathbb{R}^3 \right)^N / \mathrm{E}^+ \left( 3 \right)
\end{math}
from the RRM of the molecule in 
\begin{math}
\left( \mathbb{R}^3 \right)^N / \left( \mathrm{E} \left( 3 \right) \times \mathrm{Sym}_\Omega \right)
\end{math}
and the set of permutations occuring as the molecule moves along each reaction path in the RRM. For instance, in case of GRRM, the information on an RRM in 
\begin{math}
\left( \mathbb{R}^3 \right)^N / \left( \mathrm{E} \left( 3 \right) \times \mathrm{Sym}_\Omega \right)
\end{math}
can be obtained from the log files \verb|*_EQ_list.log| and \verb|*_TS_list.log| and information on permutation occuring  along each reaction path can be obtained from the log files \verb|*_TS*.log| \cite{afirweb}. Herein, although the proposed algorithm is demonstrated only in the case of GRRM, in principle, it should work with other potential search algorithms. We denote 
\begin{math}
\left[ r \right]
\end{math}
for the class in the shape space
\begin{math}
\left( \mathbb{R}^3 \right)^N / \mathrm{E}^+ \left( 3 \right)
\end{math}
represented by 
\begin{math}
r \in \left( \mathbb{R}^3 \right)^N
\end{math}.

\subsection{Mathematical Preliminaries}
This section reviews the mathematical concepts used in the proposed algorithm.

For a given configuration
\begin{math}
r \in \left( \mathbb{R}^3 \right)^N
\end{math}, we define the subgroup of 
\begin{math}
\mathrm{Sym}_\Omega
\end{math}
as
\begin{equation}
\mathfrak{U} \left( r \right) = \left\{ \sigma \in \mathrm{Sym}_\Omega \middle| \exists g \in \mathrm{E}^+ \left( 3 \right), \sigma \cdot r = g \cdot r \right\}.
\end{equation}
The order of 
\begin{math}
\mathfrak{U} \left( r \right)
\end{math}, 
\begin{math}
\left| \mathfrak{U} \left( r \right) \right|
\end{math}, is known as symmetry number \cite{EhrenfestDeductionOT}. For any 
\begin{math}
\sigma \in \mathrm{Sym}_\Omega
\end{math}
and 
\begin{math}
g \in \mathrm{E}^+ \left( 3 \right)
\end{math}, 
\begin{math} 
\sigma \mathfrak{U} \left( r \right) \sigma^{-1} = \mathfrak{U} \left( \sigma \cdot r \right)
\end{math}
and 
\begin{math}
\mathfrak{U} \left( g \cdot r \right) = \mathfrak{U} \left( r \right)
\end{math}
hold. The former equation holds since 
\begin{math}
\sigma' \in \mathfrak{U} \left( \sigma \cdot r \right)
\end{math}
if and only if there exists 
\begin{math}
g \in \mathrm{E}^+ \left( 3 \right)
\end{math}
such that
\begin{math}
g \cdot \sigma \cdot r = \sigma' \cdot \sigma \cdot r
\end{math}
holds, which is equivalent to 
\begin{math}
\sigma^{-1} \cdot \sigma' \sigma \cdot r = g \cdot r
\end{math}, that is,
\begin{math}
\sigma^{-1} \cdot \sigma' \sigma \in \mathfrak{U} \left( r \right)
\end{math}.
The latter equation implies that 
\begin{math}
\mathfrak{U}
\end{math}
is 
\begin{math}
\mathrm{E}^+ \left( 3 \right)
\end{math}-invariant, and thus the subgroup 
\begin{math}
\mathfrak{U} \left( r \right)
\end{math}
depends only on 
\begin{math}
\left[ r \right] \in \left( \mathbb{R}^3 \right)^N / \mathrm{E}^+ \left( 3 \right)
\end{math}. Therefore, we also write 
\begin{math}
\mathfrak{U} \left( \left[ r \right] \right)
\end{math}. In addition, the symmetry number does not depend on the choice of a representative 
\begin{math}
r \in \left( \mathbb{R}^3 \right)^N
\end{math}
of a class in 
\begin{math}
\left( \mathbb{R}^3 \right)^N / \left(\mathrm{E}^+ \left( 3 \right) \times \mathrm{Sym}_\Omega\right)
\end{math}, since 
\begin{equation}
\left| \mathfrak{U} \left( \sigma \cdot g \cdot r \right) \right| = \left| \mathfrak{U} \left( \sigma \cdot r \right) \right| = \left| \sigma \mathfrak{U} \left( r \right) \sigma^{-1} \right| = \left| \mathfrak{U} \left( r \right) \right|
\end{equation}
holds true as indicated in Ref.~\citenum{doi:10.1021/ja00478a009}. Therefore, we write the symmetry number of 
\begin{math}
r
\end{math}
as 
\begin{math}
\sigma_{\left[ \left[ r \right] \right]}
\end{math}
where 
\begin{math}
\left[ \left[ r \right] \right]
\end{math}
is the class in 
\begin{math}
\left( \mathbb{R}^3 \right)^N / \left(\mathrm{E}^+ \left( 3 \right) \times \mathrm{Sym}_\Omega\right)
\end{math}
represented by 
\begin{math}
r
\end{math}.

Using the subgroup 
\begin{math}
\mathfrak{U} \left( r \right)
\end{math}, we obtain the left coset decomposition of 
\begin{math}
\mathrm{Sym}_\Omega
\end{math}
as
\begin{equation}
\mathrm{Sym}_\Omega = \sigma_1 \mathfrak{U} \left( r \right) \cup \sigma_2 \mathfrak{U} \left( r \right) \cup \cdots \cup \sigma_m \mathfrak{U} \left( r \right), \label{eq:lcd}
\end{equation}
where 
\begin{math}
\sigma_1
\end{math}
is the identity element and 
\begin{math}
m = \left[ \mathrm{Sym}_\Omega : \mathfrak{U} \left( r \right) \right]
\end{math}.
In this case, note that 
\begin{math}
\sigma_1 \cdot r, \sigma_2 \cdot r, \sigma_3 \cdot r, \cdots, \sigma_m \cdot r
\end{math}
are permutation isomers belonging to the distinct classes in 
\begin{math}
\left( \mathbb{R}^3 \right)^N / \mathrm{E}^+ \left( 3 \right)
\end{math}.
This is because if there exists 
\begin{math}
g \in \mathrm{E}^+ \left( 3 \right)
\end{math}
such that 
\begin{math}
g \cdot \sigma_i \cdot r = \sigma_j \cdot r
\end{math}
holds for 
\begin{math}
i, j \in \left\{ 1, \cdots, m \right\}, i \neq j
\end{math}, 
\begin{math}
g \cdot r = \left( \sigma_i^{-1} \sigma_j \right) \cdot r
\end{math}
holds and 
\begin{math}
\sigma_i^{-1} \sigma_j \in \mathfrak{U} \left( r \right)
\end{math}
by definition. This contradicts the fact that 
\begin{math}
\sigma_i \mathfrak{U} \left( r \right)
\end{math}
and 
\begin{math}
\sigma_j \mathfrak{U} \left( r \right)
\end{math}
are the two distinct left cosets. Therefore, a one-to-one correspondence exists between the set of the left cosets of 
\begin{math}
\mathfrak{U} \left( r \right)
\end{math}
and the set 
\begin{equation}
\left\{ \left[ \sigma_1 \cdot r \right], \left[ \sigma_2 \cdot r \right], \cdots, \left[ \sigma_m \cdot r \right] \right\}
\end{equation}
by 
\begin{math}
\sigma_j \mathfrak{U} \left( r \right) \mapsto \left[ \sigma_j \cdot r \right]
\end{math}
for 
\begin{math}
j \in \left\{ 1, \cdots, m \right\}
\end{math}. Note that the correspondence depends on the chosen representative of 
\begin{math}
r
\end{math}
in 
\begin{math}
\left( \mathbb{R}^3 \right)^N / \left( \mathrm{E}^+ \left( 3 \right) \times \mathrm{Sym}_\Omega \right)
\end{math}. If a different representative 
\begin{math}
\tilde{r}
\end{math}
is chosen, the correspondence is as follows: Since 
\begin{math}
r
\end{math}
and 
\begin{math}
\tilde{r}
\end{math}
belong to the same class, there exist
\begin{math}
g \in \mathrm{E}^+ \left( 3 \right)
\end{math}
and
\begin{math}
\sigma \in \mathrm{Sym}_\Omega
\end{math}
such that 
\begin{math}
g \cdot \tilde{r} = \sigma \cdot r
\end{math}
holds. In this case, 
\begin{equation}
\mathfrak{U} \left( \tilde{r} \right) = \mathfrak{U} \left( g \cdot \tilde{r} \right) = \mathfrak{U} \left( \sigma \cdot r \right) = \sigma \mathfrak{U} \left( r \right) \sigma^{-1}
\end{equation}
holds, and thus, the subgroups 
\begin{math}
\mathfrak{U} \left( \tilde{r} \right)
\end{math}
and 
\begin{math}
\mathfrak{U} \left( r \right)
\end{math}
are conjugate with each other. Therefore, if Eq.~\eqref{eq:lcd} is the left coset decomposition of
\begin{math}
\mathfrak{U} \left( r \right)
\end{math}, 
\begin{equation}
\mathrm{Sym}_\Omega = \left( \sigma_1 \sigma \right) \mathfrak{U} \left( \tilde{r} \right) \cup \left( \sigma_2 \sigma \right) \mathfrak{U} \left( \tilde{r} \right) \cup \cdots \cup \left( \sigma_m  \sigma \right) \mathfrak{U} \left( \tilde{r} \right)
\end{equation}
is the left coset decomposition of 
\begin{math}
\mathfrak{U} \left( \tilde{r} \right)
\end{math}. Since 
\begin{math}
\left[ \left( \sigma_j \sigma \right) \cdot \tilde{r} \right] = \left[ \sigma_j \cdot r \right]
\end{math}
holds for all 
\begin{math}
j \in \left\{ 1, \cdots, m \right\}
\end{math}, the correspondence 
\begin{math}
\left( \sigma_j \sigma \right) \mathfrak{U} \left( \tilde{r} \right) \mapsto \left[ \left( \sigma_j \sigma \right) \cdot \tilde{r} \right]
\end{math}
provides a one-to-one correspondence between the cosets of 
\begin{math}
\mathfrak{U} \left( \tilde{r} \right)
\end{math}
and the set 
\begin{equation}
\left\{ \left[ \left( \sigma_1 \sigma \right) \cdot \tilde{r} \right], \left[ \left( \sigma_2 \sigma \right) \cdot \tilde{r} \right], \cdots, \left[ \left( \sigma_m \sigma \right) \cdot \tilde{r} \right] \right\} = \left\{ \left[ \sigma_1 \cdot r \right], \left[ \sigma_2 \cdot r \right], \cdots, \left[ \sigma_m \cdot r \right] \right\}.
\end{equation}

\begin{theorem} \label{thm:npi}
For a given configuration
\begin{math}
r \in \left( \mathbb{R}^3 \right)^N
\end{math}, 
\begin{math}
\left[ \mathrm{Sym}_\Omega : \mathfrak{U} \left( r \right) \right] \: \left( := m \right)
\end{math}
is the number of distinct permutation isomers of 
\begin{math}
r
\end{math}
in 
\begin{math}
\left( \mathbb{R}^3 \right)^N / \mathrm{E}^+ \left( 3 \right)
\end{math}. If Eq.~\eqref{eq:lcd} is the left coset decomposition of
\begin{math}
\mathfrak{U} \left( r \right)
\end{math}, 
\begin{math}
\left\{ \left[ \sigma_1 \cdot r \right], \left[ \sigma_2 \cdot r \right], \cdots, \left[ \sigma_m \cdot r \right] \right\}
\end{math}
is the set of distinct classes in 
\begin{math}
\left( \mathbb{R}^3 \right)^N / \mathrm{E}^+ \left( 3 \right)
\end{math}.
\end{theorem}

Take a transition state 
\begin{math}
r^\ddag \in \left( \mathbb{R}^3 \right)^N
\end{math}. Under the current setting, the Hesse matrix 
\begin{math}
H \left( r^\ddag \right)
\end{math}
has a unique negative eigenvalue. Let 
\begin{math}
e \in \left( \mathbb{R}^3 \right)^N
\end{math}
be a unit eigenvector of 
\begin{math}
H \left( r^\ddag \right)
\end{math}
of the negative eigenvalue. Let us consider the flow of the ordinary differential equation
\begin{equation}
\frac{dr \left( s \right)}{ds} = - \nabla V \left( r \left( s \right) \right). \label{eq:ode}
\end{equation}
Then, 
\begin{math}
r^\ddag
\end{math}
is a fixed point of the flow and the linearized equation of Eq.~\eqref{eq:ode} at 
\begin{math}
r^\ddag
\end{math}
is 
\begin{equation}
\frac{dr \left( s \right)}{ds} = - H \left( r^\ddag \right) r \left( s \right).
\end{equation}
By using Theorem~1 in Ref.~\citenum{Kelley1967}, there is a unique $1$-dimensional unstable manifold $M^+$ of 
\begin{math}
r^\ddag
\end{math}
tangent to 
\begin{math}
e
\end{math}
at 
\begin{math}
r^\ddag
\end{math},
which implies that there exist solutions 
\begin{math}
\gamma_\pm
\end{math}
of Eq.~\eqref{eq:ode} such that 
\begin{equation}
\lim_{s \rightarrow -\infty} \gamma_{\pm} \left( s \right) = r^\ddag
\end{equation}
and 
\begin{equation}
\lim_{s \rightarrow -\infty} - \frac{\nabla V \left( r_{\pm} \left( s \right) \right)}{\left\| \nabla V \left( r_{\pm} \left( s \right) \right) \right\|} = \pm e
\end{equation}
hold and they are unique up to parameter translation, i.e., if 
\begin{math}
\gamma_+ \; \left( \gamma_- \right)
\end{math}
and 
\begin{math}
\gamma_+' \; \left( \gamma_-' \right)
\end{math}
are two such solutions, there exists 
\begin{math}
s_0 \in \mathbb{R}
\end{math}
such that 
\begin{equation}
\gamma_+ \left( s \right) = \gamma_+' \left( s + s_0 \right) \; \left( \gamma_- \left( s \right) = \gamma_-' \left( s + s_0 \right) \right)
\end{equation}
holds for all 
\begin{math}
s \in \mathbb{R}
\end{math}.
Let us suppose 
\begin{math}
\lim_{s \rightarrow \infty} \gamma_- \left( s \right) = r_R
\end{math}
and
\begin{math}
\lim_{s \rightarrow \infty} \gamma_+ \left( s \right) = r_P
\end{math}
and assume they are equilibrium states. In such a case, we denote 
\begin{math}
\gamma_\pm
\end{math}
as the reaction path connecting 
\begin{math}
r_R
\end{math}
and 
\begin{math}
r_P
\end{math}
through 
\begin{math}
r^\ddag
\end{math}. We call it \emph{the} reaction path since it is unique up to parameter translation.
\begin{lemma} \label{lem:sreactionpath}
If 
\begin{math}
\gamma_{\pm} \left( s \right)
\end{math}
is the reaction path connecting 
\begin{math}
r_R
\end{math}
and 
\begin{math}
r_P
\end{math}
through 
\begin{math}
r^{\ddag}
\end{math}, then, 
\begin{math}
\sigma \cdot \gamma_{\pm} \left( s \right)
\end{math}
is the reaction path connecting 
\begin{math}
\sigma \cdot r_R
\end{math}
and 
\begin{math}
\sigma \cdot r_P
\end{math}
through 
\begin{math}
\sigma \cdot r^\ddag
\end{math}.
\end{lemma}
\begin{proof}
\begin{math}
\sigma \cdot \gamma_{\pm} \left( s \right)
\end{math}
is a solution of Eq.~\eqref{eq:ode} since
\begin{equation}
\frac{d \left( \sigma \cdot \gamma_\pm \left( s \right) \right)}{ds} = \sigma \cdot \frac{d \gamma_\pm \left( s \right)}{ds} = - \sigma \cdot \nabla V \left( \gamma_\pm \left( s \right) \right) = - \nabla V \left( \sigma \cdot \gamma_\pm \left( s \right) \right).
\end{equation}
In addition, 
\begin{equation}
\lim_{s \rightarrow -\infty} \sigma \cdot \gamma_{\pm} \left( s \right) = \sigma \cdot \left( \lim_{s \rightarrow -\infty} \gamma_{\pm} \left( s \right) \right) = \sigma \cdot r^\ddag,
\end{equation}
and 
\begin{multline}
\lim_{s \rightarrow -\infty} - \frac{\nabla V \left( \sigma \cdot \gamma_{\pm} \left( s \right) \right)}{\left\| \nabla V \left( \sigma \cdot \gamma_{\pm} \left( s \right) \right) \right\|} = \lim_{s \rightarrow -\infty} - \frac{\sigma \cdot \nabla V \left( \gamma_{\pm} \left( s \right) \right)}{\left\| \sigma \cdot \nabla V \left( \gamma_{\pm} \left( s \right) \right) \right\|} \\
= \sigma \cdot \left( \lim_{s \rightarrow -\infty} - \frac{\nabla V \left( \gamma_{\pm} \left( s \right) \right)}{\left\| \nabla V \left( \gamma_{\pm} \left( s \right) \right) \right\|} \right) = \pm \sigma \cdot e
\end{multline}
hold. Since 
\begin{math}
V
\end{math}
is 
\begin{math}
\mathrm{Sym}_\Omega
\end{math}-invariant, 
\begin{math}
\sigma \cdot r_R
\end{math}
and 
\begin{math}
\sigma \cdot r_P
\end{math}
are equilibrium states, 
\begin{math}
\sigma \cdot r^\ddag
\end{math}
is a transition state, and 
\begin{math}
\sigma \cdot e
\end{math}
is a unit eigenvector corresponding to the negative eigenvalues of 
\begin{math}
H \left( \sigma \cdot r^\ddag \right)
\end{math}. This proves the lemma.
\end{proof}

By Theorem~\ref{thm:npi}, there are 
\begin{math}
m^\ddag = \left[ \mathrm{Sym}_\Omega : \mathfrak{U} \left( r^\ddag \right) \right]
\end{math}
distinct reaction paths up to the action of 
\begin{math}
\mathrm{E}^+ \left( 3 \right)
\end{math}
corresponding to 
\begin{math}
\gamma_{\pm}
\end{math}.
Suppose 
\begin{equation}
\mathrm{Sym}_\Omega = \sigma_1^\ddag \mathfrak{U} \left( r^\ddag \right) \cup \sigma_2^\ddag \mathfrak{U} \left( r^\ddag \right) \cup \cdots \cup \sigma_{m^\ddag}^\ddag \mathfrak{U} \left( r^\ddag \right)
\end{equation}
is the left coset decomposition of 
\begin{math}
\mathrm{Sym}_\Omega
\end{math}
by the subgroup 
\begin{math}
\mathfrak{U} \left( r^\ddag \right)
\end{math}. The 
\begin{math}
m^\ddag
\end{math}
reaction paths 
\begin{math}
\sigma_j^\ddag \cdot \gamma_{\pm} \left( s \right)
\end{math}
connect 
\begin{math}
\sigma_j^\ddag \cdot r_R
\end{math}
and 
\begin{math}
\sigma_j^\ddag \cdot r_P
\end{math}
through 
\begin{math}
\sigma_j^\ddag \cdot r^\ddag
\end{math}
for 
\begin{math}
j \in \left\{ 1, \cdots, m^\ddag \right\}
\end{math}
by Lemma~\ref{lem:sreactionpath}. 
\begin{lemma} \label{lem:endpoints}
In this setting, the set 
\begin{math}
\left\{ \left[ \sigma_j^\ddag \cdot r_R \right], \left[ \sigma_j^\ddag \cdot r_P \right] \right\}
\end{math}
does not depend on the choice of a representative of the coset 
\begin{math}
\sigma_j^\ddag \mathfrak{U} \left( r^\ddag \right)
\end{math}. 
\end{lemma}
\begin{proof}
Suppose 
\begin{math}
\bar{\sigma}_j^\ddag = \sigma_j^\ddag \sigma
\end{math}
for a 
\begin{math}
\sigma \in \mathfrak{U} \left( r^\ddag \right)
\end{math}. By the definition, there exists 
\begin{math}
\left( t, g \right) \in \mathrm{E}^+ \left( 3 \right)
\end{math}
such that 
\begin{math}
\sigma \cdot r^\ddag = \left( t, g \right) \cdot r^\ddag
\end{math}
holds. Without loss of generality, we can assume 
\begin{math}
t = 0
\end{math}
since the coordinate in 
\begin{math}
\left( \mathbb{R}^3 \right)^N
\end{math}
can be chosen so that the center of the mass of 
\begin{math}
r^\ddag
\end{math}
is the origin and the center of mass is invariant by the action of $\sigma$. By the property of the Hesse matrix, 
\begin{math}
H \left( \sigma \cdot r^\ddag \right) \; \left( = H \left( g \cdot r^\ddag \right) \right)
\end{math}
has one negative eigenvalue and its eigenvector is 
\begin{math}
\sigma \cdot e \; \left( g \cdot e \right)
\end{math}. Since 
\begin{math}
\sigma \cdot e
\end{math}
and 
\begin{math}
g \cdot e
\end{math}
have the same length, there can be two possibilities: 
\begin{enumerate}
\item
\begin{math}
\sigma \cdot e = g \cdot e
\end{math},
\item
\begin{math}
\sigma \cdot e = - g \cdot e
\end{math}.
\end{enumerate}
In the first case, since both
\begin{math}
g \cdot \gamma_\pm \left( s \right)
\end{math}
and 
\begin{math}
\sigma \cdot \gamma_\pm \left( s \right)
\end{math}
satisfy Eq.~\eqref{eq:ode} and are asymptotic to 
\begin{math}
\sigma \cdot r^\ddag = g \cdot r^\ddag
\end{math}
in the direction 
\begin{math}
\pm g \cdot e \; \left( = \pm \sigma \cdot e \right)
\end{math}, the uniqueness guarantees that 
\begin{math}
g \cdot \gamma_\pm \left( s \right)
\end{math}
and 
\begin{math}
\sigma \cdot \gamma_\pm \left( s \right)
\end{math}
coincide up to parameter translation. This implies that 
\begin{math}
g \cdot r_R = \sigma \cdot r_R
\end{math}
and 
\begin{math}
g \cdot r_P = \sigma \cdot r_P
\end{math}
by taking the limit 
\begin{math}
s \rightarrow \infty
\end{math}. Using this, we obtain 
\begin{math}
\bar{\sigma}_j^\ddag \cdot r_R = \left( \sigma_j^\ddag \sigma \right) \cdot r_R = \sigma_j^\ddag \cdot \left( g \cdot r_R \right) = g \cdot \left( \sigma_j^\ddag \cdot r_R \right)
\end{math}
and 
\begin{math}
\bar{\sigma}_j^\ddag \cdot r_P = \left( \sigma_j^\ddag \sigma \right) \cdot r_P = \sigma_j^\ddag \cdot \left( g \cdot r_P \right) = g \cdot \left( \sigma_j^\ddag \cdot r_P \right)
\end{math}
and thus 
\begin{math}
\left[ \bar{\sigma}_j^\ddag \cdot r_R \right] = \left[ \sigma_j^\ddag \cdot r_R \right]
\end{math}
and 
\begin{math}
\left[ \bar{\sigma}_j^\ddag \cdot r \right] = \left[ \sigma_j^\ddag \cdot r \right]
\end{math}
hold. 

\noindent
In the second case, both 
\begin{math}
g \cdot \gamma_\pm \left( s \right)
\end{math}
and 
\begin{math}
\sigma \cdot \gamma_\mp \left( s \right)
\end{math}
satisfy Eq.~\eqref{eq:ode} and are asymptotic to 
\begin{math}
\sigma \cdot r^\ddag = g \cdot r^\ddag
\end{math}
in the direction 
\begin{math}
\pm g \cdot e \; \left( = \mp \sigma \cdot e \right)
\end{math}, the uniqueness guarantees that 
\begin{math}
g \cdot \gamma_\pm \left( s \right)
\end{math}
and 
\begin{math}
\sigma \cdot \gamma_\mp \left( s \right)
\end{math}
coincide up to parameter translation. This implies that 
\begin{math}
g \cdot r_R = \sigma \cdot r_P
\end{math}
and 
\begin{math}
g \cdot r_P = \sigma \cdot r_R
\end{math}
by taking the limit 
\begin{math}
s \rightarrow \infty
\end{math}. In this case, the reactant and product are permutation isomers. In this case, we obtain 
\begin{math}
\bar{\sigma}_j^\ddag \cdot r_R = \left( \sigma_j^\ddag \sigma \right) \cdot r_R = \sigma_j^\ddag \cdot \left( g \cdot r_P \right) = g \cdot \left( \sigma_j^\ddag \cdot r_P \right)
\end{math}
and 
\begin{math}
\bar{\sigma}_j^\ddag \cdot r_P = \left( \sigma_j^\ddag \sigma \right) \cdot r_P = \sigma_j^\ddag \cdot \left( g \cdot r_R \right) = g \cdot \left( \sigma_j^\ddag \cdot r_R \right)
\end{math}
and thus 
\begin{math}
\left[ \bar{\sigma}_j^\ddag \cdot r_R \right] = \left[ \sigma_j^\ddag \cdot r_P \right]
\end{math}
and 
\begin{math}
\left[ \bar{\sigma}_j^\ddag \cdot r_P \right] = \left[ \sigma_j^\ddag \cdot r_R \right]
\end{math}
hold. In the both cases, the set 
\begin{math}
\left\{ \left[ \sigma_j^\ddag \cdot r_R \right], \left[ \sigma_j^\ddag \cdot r_P \right] \right\}
\end{math}
does not depend on the choice of a representative of the coset 
\begin{math}
\sigma_j^\ddag \mathfrak{U} \left( r^\ddag \right)
\end{math}. 
\end{proof}
Using Lemma~\ref{lem:endpoints}, we present an algorithm to reproduce the RRM in the shape space in what follows. We identify RRM as a multi-graph 
\begin{math}
\mathrm{G} = \left( \mathrm{V}, \mathrm{E}, h, \pi_{\mathrm{V}}, \pi_{\mathrm{E}} \right)
\end{math}
where 
\begin{math}
\mathrm{V}
\end{math}
is the set of vertices consisting of distinct minima of the potential energy function in the shape space and 
\begin{math}
\mathrm{E}
\end{math}
is the set of edges consisting of distinct reaction paths in the shape space connecting minima. We identify 
\begin{math}
\mathrm{E}
\end{math}
to the set of distinct transition states in the shape space 
\begin{math}
\left( \mathbb{R}^3 \right)^N / \mathrm{E}^+ \left( 3 \right)
\end{math}
since there is a one-to-one correspondence between the set of reaction paths and transition states. 
\begin{math}
h \colon \mathrm{E} \rightarrow \left\{ \left\{ v, w \right\} \middle| v, w \in V \right\}
\end{math}
is the map assigning to each edge the set of its endpoint vertices, 
\begin{math}
\pi_{\mathrm{V}} \colon \mathrm{V} \rightarrow \left( \mathbb{R}^3 \right)^N / \mathrm{E}^+ \left( 3 \right) \times \mathrm{Sym}_\Omega
\end{math}
and
\begin{math}
\pi_{\mathrm{E}} \colon \mathrm{E} \rightarrow \left( \mathbb{R}^3 \right)^N / \mathrm{E}^+ \left( 3 \right) \times \mathrm{Sym}_\Omega
\end{math}
are the projection from the shape space 
\begin{math}
\left( \mathbb{R}^3 \right)^N / \mathrm{E}^+ \left( 3 \right)
\end{math}
to 
\begin{math}
\left( \mathbb{R}^3 \right)^N / \mathrm{E}^+ \left( 3 \right) \times \mathrm{Sym}_\Omega
\end{math}. We refer other possible formulations of RRMs to Ref.~\citenum{Temkin}.
\subsection{Algorithm to reproduce RRM on the shape space} \label{sec:algrep}
The proposed algorithm to reproduce RRM in the shape space
\begin{math}
\left( \mathbb{R}^3 \right)^N / \mathrm{E}^+ \left( 3 \right)
\end{math}
from the list of tuples of transition state configuration 
\begin{math}
r_T
\end{math} and the set of reactant and product configurations 
\begin{math}
\left\{ r_R, r_P \right\}
\end{math} is as follows: Suppose 
\begin{math}
\left\{ \left( r_T^{\left( i \right)}, \left\{ r_R^{\left( i \right)}, r_P^{\left( i \right)} \right\} \right) \right\}_{i \in \tilde{I}}
\end{math}
is the list of tuples of transition state configuration 
\begin{math}
r_T
\end{math} and the set of reactant and product configurations 
\begin{math}
\left\{ r_R, r_P \right\}
\end{math}
and 
\begin{math}
\tilde{I}
\end{math}
is the index set obtained using a potential search algorithm such as GRRM and 
\begin{math}
\left\{ r_{\mathrm{EQ}}^{\left( j \right)} \right\}_{j \in \tilde{J}}
\end{math}
is the list of the equilibrium state configurations distinct up to the action of
\begin{math}
\mathrm{E} \left( 3 \right) \times \mathrm{Sym}_\Omega
\end{math}
where 
\begin{math}
\tilde{J}
\end{math}
is the index set of the list. For 
\begin{math}
j \in \tilde{J}
\end{math}, if
\begin{math}
r_{\mathrm{EQ}}^{\left( j \right)}
\end{math}
and 
\begin{math}
\mathrm{i} \cdot r_{\mathrm{EQ}}^{\left( j \right)}
\end{math}
belong to two different orbits of the action 
\begin{math}
\mathrm{E}^+ \left( 3 \right) \times \mathrm{Sym}_\Omega
\end{math}, i.e., they are chiral, consider
\begin{math}
r_{\mathrm{EQ}}^{\left( j \right)}
\end{math}
and 
\begin{math}
\mathrm{i} \cdot r_{\mathrm{EQ}}^{\left( j \right)}
\end{math}
as the two distinct elements. Similarly for 
\begin{math}
i \in \tilde{I}
\end{math}
if 
\begin{math}
r_{\mathrm{T}}^{\left( i \right)}
\end{math}
and 
\begin{math}
\mathrm{i} \cdot r_{\mathrm{T}}^{\left( i \right)}
\end{math}
belong to two different orbits of the action 
\begin{math}
\mathrm{E}^+ \left( 3 \right) \times \mathrm{Sym}_\Omega
\end{math}, consider
\begin{math}
r_{\mathrm{EQ}}^{\left( i \right)}
\end{math}
and 
\begin{math}
\mathrm{i} \cdot r_{\mathrm{T}}^{\left( i \right)}
\end{math}
as the two distinct elements and add 
\begin{math}
\left( \mathrm{i} \cdot r_T^{\left( i \right)}, \left\{ \mathrm{i} \cdot r_R^{\left( i \right)}, \mathrm{i} \cdot r_P^{\left( i \right)} \right\} \right)
\end{math}
to the list. We redefine the resulting lists as 
\begin{math}
\left\{ r_{\mathrm{EQ}}^{\left( j \right)} \right\}_{j \in J}
\end{math}
and
\begin{math}
\left\{ \left( r_T^{\left( i \right)}, \left\{ r_R^{\left( i \right)}, r_P^{\left( i \right)} \right\} \right) \right\}_{i \in I}
\end{math}.
\begin{enumerate}
\item Initiate 
\begin{math}
\mathrm{E} = \emptyset
\end{math}
and 
\begin{math}
h \colon \emptyset \rightarrow \left\{ \left\{ v, w \right\} \middle| v, w \in \mathrm{V} \right\}
\end{math}.
\item For each 
\begin{math}
j \in J
\end{math}, compute the left coset decomposition of 
\begin{math}
\mathrm{Sym}_\Omega
\end{math}
by 
\begin{math}
\mathfrak{U} \left( r_{\mathrm{EQ}}^{\left( j \right)} \right)
\end{math}, i.e., 
\begin{equation}
\mathrm{Sym}_\Omega = \sigma_1^{\left( j \right)} \mathfrak{U} \left( r_{\mathrm{EQ}}^{\left( j \right)} \right) \cup \sigma_2^{\left( j \right)} \mathfrak{U} \left( r_{\mathrm{EQ}}^{\left( j \right)} \right) \cup \cdots \cup \sigma_{m^{\left( j \right)}}^{\left( j \right)} \mathfrak{U} \left( r_{\mathrm{EQ}}^{\left( j \right)} \right). 
\end{equation}
\item Set the vertex set as
\begin{equation}
\mathrm{V} = \bigcup_{j \in J} \left\{ \left[ \sigma_1^{\left( j \right)} \cdot r_{\mathrm{EQ}}^{\left( j \right)} \right], \left[ \sigma_2^{\left( j \right)} \cdot r_{\mathrm{EQ}}^{\left( j \right)} \right], \cdots, \left[ \sigma_{m^{\left( j \right)}}^{\left( j \right)} \cdot r_{\mathrm{EQ}}^{\left( j \right)} \right] \right\}
\end{equation}
\item For each 
\begin{math}
i \in I
\end{math}, compute the left coset decomposition of 
\begin{math}
\mathrm{Sym}_\Omega
\end{math}
by 
\begin{math}
\mathfrak{U} \left( r_T^{\left( i \right)} \right)
\end{math}, i.e., 
\begin{equation}
\mathrm{Sym}_\Omega = \sigma_1^{\left( i \right), \ddag} \mathfrak{U} \left( r_T^{\left( i \right)} \right) \cup \sigma_2^{\left( i \right), \ddag} \mathfrak{U} \left( r_T^{\left( i \right)} \right) \cup \cdots \cup \sigma_{m^{\left( i \right),\ddag}}^{\left( i \right), \ddag} \mathfrak{U} \left( r_T^{\left( i \right)} \right). 
\end{equation}
\item Add an edge 
\begin{math}
\left[ \sigma_k^{\left( i \right), \ddag} r_T^{\left( i \right)} \right]
\end{math}
to 
\begin{math}
\mathrm{E}
\end{math}
and set 
\begin{math}
h \left( \left[ \sigma_k^{\left( i \right), \ddag} r_T^{\left( i \right)} \right] \right) = \left\{ \left[ \sigma_k^{\left( i \right), \ddag} \cdot r_R^{\left( i \right)} \right], \left[ \sigma_k^{\left( i \right), \ddag} \cdot r_P^{\left( i \right)} \right] \right\}
\end{math}
for all 
\begin{math}
k \in \left\{ 1, \cdots, m^{\left( i \right),\ddag} \right\}
\end{math}.
\end{enumerate}
Note that each edge in 4. does not depend on the chosen representative 
\begin{math}
\sigma_k^{\left( i \right), \ddag}
\end{math}
in the coset 
\begin{math}
\sigma_k^{\left( i \right), \ddag} \mathfrak{U} \left( r_T^{\left( i \right)} \right)
\end{math}. Identification of  
\begin{math}
\left[ \sigma_k^{\left( i \right), \ddag} \cdot r_R^{\left( i \right)} \right]
\end{math}
(\begin{math}
\left[ \sigma_k^{\left( i \right), \ddag} \cdot r_P^{\left( i \right)} \right]
\end{math}) to one of the left cosets of 
\begin{math}
\mathfrak{U}
\end{math}
of the equilibrium structure,
\begin{math}
r_{\mathrm{EQ}}^{\left( j_{R \left( P \right)} \right)}
\end{math}, can be performed as follows: First, find 
\begin{math}
j_R \in J
\end{math},
\begin{math}
\sigma \in \mathrm{Sym}_\Omega
\end{math}
and 
\begin{math}
g \in \mathrm{E}^+ \left( 3 \right)
\end{math}
such that 
\begin{math}
g \cdot r_R^{\left( i \right)} = \sigma \cdot r_{\mathrm{EQ}}^{\left( j_{R} \right)}
\end{math}
holds true. Next, find the left coset of 
\begin{math}
\mathfrak{U} \left( r_{\mathrm{EQ}}^{\left( j_{R} \right)} \right)
\end{math}
containing 
\begin{math}
\sigma_k^{\left( i \right), \ddag} \cdot \sigma
\end{math}. If 
\begin{math}
\sigma_k^{\left( i \right), \ddag} \cdot \sigma \in \sigma^{\left( j_{R} \right)}_\ell \mathfrak{U} \left( r_{\mathrm{EQ}}^{\left( j_{R} \right)} \right)
\end{math}
holds, then, 
\begin{math}
\left[ \sigma_k^{\left( i \right), \ddag} \cdot r_R^{\left( i \right)} \right] = \left[ \sigma^{\left( j_R \right)}_\ell \cdot r_{\mathrm{EQ}}^{\left( j_{R} \right)} \right]
\end{math}
holds. The same is true for the product. Then, the graph 
\begin{math}
\mathrm{G} = \left( \mathrm{V}, \mathrm{E}, h, \pi_{\mathrm{V}}, \pi_{\mathrm{E}}  \right)
\end{math}
is the RRM in the space
\begin{math}
\left( \mathbb{R}^3 \right)^N / \mathrm{E}^+ \left( 3 \right)
\end{math}
corresponding to the input of the list of tuples obtained by using potential search algorithms such as GRRM.

\section{Isomorphisms between Subgraphs mapped with each other by the action of $\mathrm{Sym}_\Omega$} \label{sec:graphisomorphism}
Sometimes the resulting graph 
\begin{math}
\mathrm{G}
\end{math}
may have several connected components mapped to each other by the action of 
\begin{math}
\mathrm{Sym}_\Omega
\end{math}.
In this case, all such connected components are isomorphic. To formulate this, let us define two multi-graphs in this context are isomorphic.
\begin{definition}[graph isomorphism] \label{def;graph_isomorphism}
Two graphs 
\begin{math}
\mathrm{G} = \left( \mathrm{V}, \mathrm{E}, h, \pi_{\mathrm{V}}, \pi_{\mathrm{E}}  \right)
\end{math}
and 
\begin{math}
\mathrm{G}' = \left( \mathrm{V}', \mathrm{E}', h', \pi'_{\mathrm{V}'}, \pi'_{\mathrm{E}'}  \right)
\end{math}
are isomorphic if there exist bijections 
\begin{math}
\phi \colon \mathrm{V} \rightarrow \mathrm{V}'
\end{math}
and 
\begin{math}
\psi \colon \mathrm{E} \rightarrow \mathrm{E}'
\end{math}
such that the following diagram commutes:
\begin{equation}
  \begin{CD}
     \mathrm{V} @>{\pi_{\mathrm{V}}}>> \left( \mathbb{R}^3 \right)^N / \mathrm{E}^+ \left( 3 \right) \times \mathrm{Sym}_\Omega \\
  @V{\phi}VV    @| \\
     \mathrm{V}' @>{\pi'_{\mathrm{V}'}}>>\left( \mathbb{R}^3 \right)^N / \mathrm{E}^+ \left( 3 \right) \times \mathrm{Sym}_\Omega
  \end{CD}
\end{equation}
\begin{equation}
  \begin{CD}
     2^{\mathrm{V}} @<{h}<< \mathrm{E} @>{\pi_{\mathrm{E}}}>> \left( \mathbb{R}^3 \right)^N / \mathrm{E}^+ \left( 3 \right) \times \mathrm{Sym}_\Omega \\
  @V{\phi}VV  @V{\psi}VV  @| \\
     2^{\mathrm{V}'} @<{h'}<< \mathrm{E}' @>{\pi'_{\mathrm{E}'}}>>\left( \mathbb{R}^3 \right)^N / \mathrm{E}^+ \left( 3 \right) \times \mathrm{Sym}_\Omega
  \end{CD}
\end{equation}
where 
\begin{math}
\phi \colon 2^{\mathrm{V}} \rightarrow 2^{\mathrm{V}'}
\end{math}
is the set--valued function induced from 
\begin{math}
\phi
\end{math}. 
\end{definition}
Take an arbitrary
\begin{math}
\sigma \in \mathrm{Sym}_\Omega
\end{math}. Let 
\begin{math}
\sigma
\end{math}
act on 
\begin{math}
\mathrm{V}
\end{math}
as 
\begin{equation}
\sigma \cdot \left[ \sigma_l^{\left( j \right)} \cdot r^{\left( j \right)}_{\mathrm{EQ}} \right] = \left[ \sigma \cdot \sigma_l^{\left( j \right)} \cdot r^{\left( j \right)}_{\mathrm{EQ}} \right]
\end{equation}
for 
\begin{math}
j \in J
\end{math}
and 
\begin{math}
l \in \left\{ 1, \cdots, m^{\left( j \right)} \right\}
\end{math}
and act on 
\begin{math}
\mathrm{E}
\end{math}
as
\begin{equation}
\sigma \cdot \left[ \sigma_l^{\left( i \right)} \cdot r^{\left( i \right)}_{\mathrm{T}} \right] = \left[ \sigma \cdot \sigma_l^{\left( i \right)} \cdot r^{\left( i \right)}_{\mathrm{T}} \right]
\end{equation}
for 
\begin{math}
i \in I
\end{math}
and 
\begin{math}
l \in \left\{ 1, \cdots, m^{\left( i \right), \ddag} \right\}
\end{math}. The action is well-defined since the action of 
\begin{math}
\mathrm{Sym}_\Omega
\end{math}
and 
\begin{math}
\mathrm{E}^+ \left( 3 \right)
\end{math}
commutes. Under the action, 
\begin{math}
\pi_{\mathrm{V}}, \pi_{\mathrm{E}}
\end{math}
are
\begin{math}
\mathrm{Sym}_\Omega
\end{math}-invariant, i.e., 
\begin{math}
\pi_{\mathrm{V}} \left( \sigma \cdot v \right) = \pi_{\mathrm{V}} \left( v \right)
\end{math}
for 
\begin{math}
v \in \mathrm{V}
\end{math}
and
\begin{math}
\pi_{\mathrm{E}} \left( \sigma \cdot e \right) = \pi_{\mathrm{E}} \left( e \right)
\end{math}
for 
\begin{math}
e \in \mathrm{E}
\end{math}.
Under the action, 
\begin{math}
h
\end{math}
is 
\begin{math}
\mathrm{Sym}_\Omega
\end{math}-equivariant, i.e. 
\begin{math}
h \left( \sigma \cdot e \right) = \sigma \cdot h \left( e \right)
\end{math}, where 
\begin{math}
\sigma
\end{math}
is supposed to act on 
\begin{math}
2^\mathrm{V}
\end{math}
in the element-wise manner, since if 
\begin{math}
\gamma_{\pm}
\end{math}
is the reaction path connecting 
\begin{math}
r_R
\end{math}
and 
\begin{math}
r_P
\end{math}
through 
\begin{math}
r^\ddag
\end{math}, then, 
\begin{math}
\sigma \cdot \gamma_{\pm}
\end{math}
is the reaction path connecting 
\begin{math}
\sigma \cdot r_R
\end{math}
and 
\begin{math}
\sigma \cdot r_P
\end{math}
through 
\begin{math}
\sigma \cdot r^\ddag
\end{math}
by Lemma~\ref{lem:sreactionpath}. This implies that 
\begin{equation}
h \left( \sigma \cdot \left[ r^\ddag \right] \right) = \left\{ \left[ \sigma \cdot r_R \right], \left[ \sigma \cdot r_P \right] \right\} = \sigma \cdot \left\{ \left[ r_R \right], \left[ r_P \right] \right\} = \sigma \cdot h \left( \left[ r^\ddag \right] \right)
\end{equation}
holds. This proves the claim.

Let 
\begin{math}
\check{\mathrm{G}} = \left( \check{\mathrm{V}}, \check{\mathrm{E}}, \left. h \right|_{\check{\mathrm{E}}}, \left. \pi_{\mathrm{V}} \right|_{\check{\mathrm{V}}}, \left. \pi_{\mathrm{E}} \right|_{\check{\mathrm{E}}} \right)
\end{math}
be a subgraph of 
\begin{math}
\mathrm{G}
\end{math}, i.e. 
\begin{math}
\check{\mathrm{V}} \subset \mathrm{V}
\end{math}, 
\begin{math}
\check{\mathrm{E}} \subset \mathrm{E}
\end{math}, and 
\begin{math}
h \left( e \right) \subset \check{\mathrm{V}}
\end{math}
for all 
\begin{math}
e \in \check{\mathrm{E}}
\end{math}. Then, the action induces the action of 
\begin{math}
\mathrm{Sym}_\Omega
\end{math}
to the set of the subgraphs of 
\begin{math}
G
\end{math}
as
\begin{multline}
\sigma \cdot \check{G} = \left( \sigma \cdot \check{\mathrm{V}}, \sigma \cdot \check{\mathrm{E}}, \left. \sigma \cdot h \cdot \sigma^{-1} \right|_{\sigma \cdot \check{\mathrm{E}}}, \left. \pi_{\mathrm{V}} \cdot \sigma^{-1} \right|_{\sigma \cdot \check{\mathrm{V}}}, \left. \pi_{\mathrm{E}} \cdot \sigma^{-1} \right|_{\sigma \cdot \check{\mathrm{E}}} \right) \\
= \left( \sigma \cdot \check{\mathrm{V}}, \sigma \cdot \check{\mathrm{E}}, \left. h \right|_{\sigma \cdot \check{\mathrm{E}}}, \left. \pi_{\mathrm{V}} \right|_{\sigma \cdot \check{\mathrm{V}}}, \left. \pi_{\mathrm{E}} \right|_{\sigma \cdot \check{\mathrm{E}}} \right).
\end{multline}
Then, 
\begin{math}
\check{\mathrm{G}}
\end{math}
and 
\begin{math}
\sigma \cdot \check{\mathrm{G}}
\end{math}
are isomorphic by taking 
\begin{math}
v \mapsto \sigma \cdot v
\end{math}, 
\begin{math}
e \mapsto \sigma \cdot e
\end{math}
as 
\begin{math}
\phi, \psi
\end{math}, respectively, in Definition~\ref{def;graph_isomorphism}. Therefore, we obtain the following theorem.
\begin{theorem}
Let
\begin{math}
\check{G}_1
\end{math}
and 
\begin{math}
\check{G}_2
\end{math}
be two subgraphs of 
\begin{math}
G
\end{math}
mapped with each other by the action of 
\begin{math}
\textnormal{Sym}_\Omega
\end{math}, i.e., 
\begin{math}
\check{G}_2 = \sigma \cdot \check{G}_1
\end{math}
for some 
\begin{math}
\sigma \in \textnormal{Sym}_\Omega
\end{math}, then, 
\begin{math}
\check{G}_1
\end{math}
and 
\begin{math}
\check{G}_2
\end{math}
are isomorphic in the sense of Definition~\ref{def;graph_isomorphism}.
\end{theorem}
Specifically, if 
\begin{math}
\check{\mathrm{G}}
\end{math}
is a connected component of 
\begin{math}
\mathrm{G}
\end{math}, then, 
\begin{math}
\sigma \cdot \check{\mathrm{G}}
\end{math}
is also a connected component of 
\begin{math}
\mathrm{G}
\end{math}.

\begin{figure}[ht]
\includegraphics[keepaspectratio, scale=0.8] {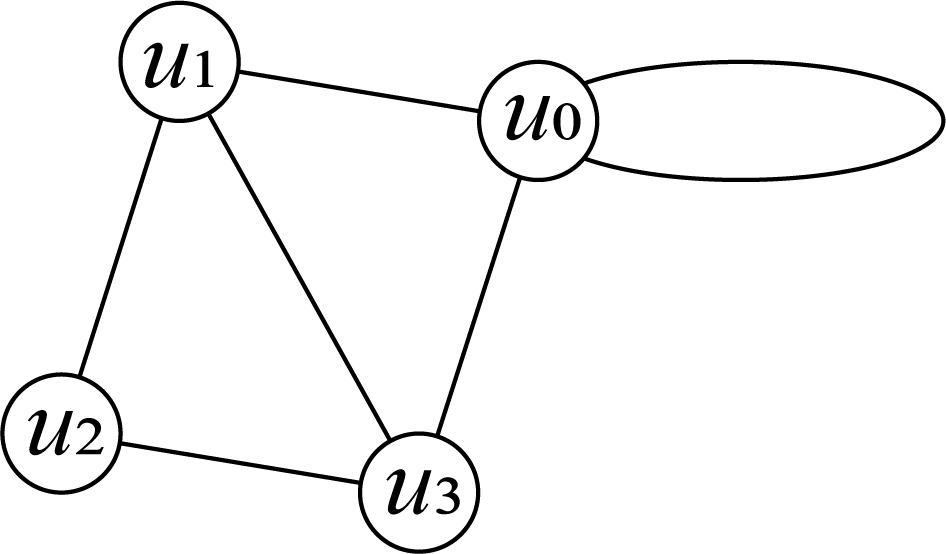}
\caption{An RRM in \begin{math}
\left( \mathbb{R}^3 \right)^N / \mathrm{E}^+ \left( 3 \right) \times \mathrm{Sym}_\Omega
\end{math} with vertices $u_0$, $u_1$, $u_2$, and $u_3$.}
\label{fig:reducedRRM}
\end{figure}

Sometimes 
\begin{math}
\mathrm{G}
\end{math}
has numerous connected components mapped to each other by the action of 
\begin{math}
\mathrm{Sym}_\Omega
\end{math}
and the computation of the entire graph 
\begin{math}
\mathrm{G}
\end{math}
is infeasible as in the case of Section~\ref{sec:demoC5H12}. In this case, all the connected components are isomorphic in the sense of Definition~ and it is enough to compute a single connected component of an entire graph. Here, we provide an algorithm to accomplish this goal. This part is a bit technical and the detail is shown in Sec.~1 in Supporting Information. Here, we provide an intuitive description of the algorithm by taking an input RRM 
$\pi\mathrm{G}_0$ in \begin{math}
\left( \mathbb{R}^3 \right)^N / \mathrm{E}^+ \left( 3 \right) \times \mathrm{Sym}_\Omega
\end{math}
in Figure.~\ref{fig:reducedRRM} as an example.
Note that the vertices correspond to equilibrium structures and edges corresponding to the reaction paths connecting them in the quotient space \begin{math}
\left( \mathbb{R}^3 \right)^N / \mathrm{E}^+ \left( 3 \right) \times \mathrm{Sym}_\Omega
\end{math}. First, we fix a root equilibrium 
\begin{math}
u_0
\end{math}, which can be arbitrary. If we continuously deform a representative conformation of 
\begin{math}
u_0
\end{math}
along a cycle 
\begin{math}
\langle u_0, u_1, u_3, u_0 \rangle
\end{math}, the conformation goes back to the original representative conformation with some of the identical atoms being permuted. Contrastingly, even if we continuously deform a representative conformation of $u_0$ along a cycle
\begin{math}
\langle u_0, u_1, u_2, u_1, u_0 \rangle
\end{math}, the resulting conformation goes back to exactly the same reference conformation. In the latter case, the cycle 
\begin{math}
\langle u_0, u_1, u_2, u_1, u_0 \rangle
\end{math}
is the path starting from $u_0$, going to $u_2$, and going back to $u_0$ along the same path and the path
can be continuously deformed to a trivial cycle, i.e., homotopic to a trivial cycle. In the latter case, the fact that the resulting conformation is exactly the same as the starting conformation, which is one of the consequences of homotopy lifting property explained in Sec.~1 in Supporting Information. Since all the cycles  are generated by fundamental cycles of the graph up to homotopy equivalence, to identify the set of non-trivial permutations occurring for reference conformations of $u_0$ by deformations along the reaction paths, it is enough to consider permutations occurring along the set of fundamental cycles starting and ending at $u_0$. In this example, fundamental cycles are
\begin{math}
\langle u_0, u_1, u_3 \rangle
\end{math}, 
\begin{math}
\langle u_0, u_1, u_2, u_3, u_0 \rangle
\end{math}, and the self-loop emanating from $u_0$. By letting 
\begin{math}
\sigma_1, \sigma_2
\end{math}, and 
\begin{math}
\sigma_3
\end{math}
be the permutations occurring by the deformations along the respective cycles, the set of permutation is generated by 
\begin{math}
\sigma_1, \sigma_2, \sigma_3
\end{math}, and 
\begin{math}
\mathfrak{U} \left( c \left( u_0 \right) \right)
\end{math}. By letting the resulting permutation group be
\begin{math}
\textnormal{Sym}_\Omega^c \left( \pi \mathrm{G}_0 \right)
\end{math}, and using it instead of 
\begin{math}
\textnormal{Sym}_\Omega
\end{math}
in Sec.~\ref{sec:algrep}, we obtain a single connected component of the RRM in the shape space corresponding to $\pi \mathrm{G}_0$. For detail, see Sec.~1 in Supporting Information.
\section{Demonstration of Algorithm in Simple Isomerization Reactions}
We demonstrate the algorithm in the previous section in simple isomerization reactions, isomerization reaction of bi-tetrahedron (trigonal-bipyramidal molecule) and Berry's pseudo rotation mechanism \cite{doi:10.1063/1.1730820}. Note that several studies have been conducted on such simple isomerization reactions and their resulting RRMs in the shape space, which sometimes called "reaction graph" in the context of chemical graph theory, starting from the work of Balaban \cite{Balaban1966}. The results in this section are by no means new. Our purpose here is to demonstrate the proposed algorithm in these simple systems before demonstrating that in more realistic, complicated systems in Sec.~\ref{sec:demo_reals}.

\subsection{Isomerization reaction of bi-tetrahedron}
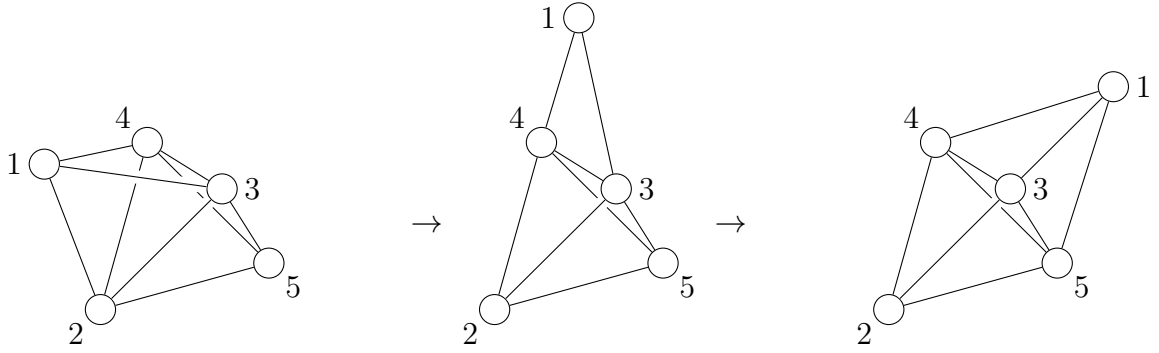
\begin{figure}[ht]
\begin{tabular}{ccc}
\begin{minipage}[b]{0.3\linewidth}
\centering
	\begin{tikzpicture}[scale=0.8]
		\node[draw, shape=circle, inner sep=4pt] (a) at (1,1,1){};
		\node[draw, shape=circle, inner sep=4pt] (b) at (-1,-1,1){};
		\node[draw, shape=circle, inner sep=4pt] (c) at (1,-1,-1){};
		\node[draw, shape=circle, inner sep=4pt] (d) at (-1,1,-1){};
		\node[draw, shape=circle, inner sep=4pt] (e) at (-5/3,5/3,5/3){};
		\node (aa) at (1.5,1,1) {$3$};\node (bb) at (-1.4,-1.4,1) {$2$};\node (cc) at (1.4,-1.4,-1) {$5$};\node (cc) at (-1.4,1.4,-1) {$4$};\node (ee) at (-13/6,5/3,5/3){$1$};
		\draw (c)--(a)--(d)--(c)--(b);\draw(b)--(d);\draw[white, line width=5pt](e)--(a)--(b);\draw(a)--(b);
		\draw (a)--(e)--(d);\draw(b)--(e);
	\end{tikzpicture}
\end{minipage}&
\begin{minipage}[b]{0.35\linewidth}
\centering
	\begin{tikzpicture}[scale=0.8]
		\node[draw, shape=circle, inner sep=4pt] (a) at (1,1,1){};
		\node[draw, shape=circle, inner sep=4pt] (b) at (-1,-1,1){};
		\node[draw, shape=circle, inner sep=4pt] (c) at (1,-1,-1){};
		\node[draw, shape=circle, inner sep=4pt] (d) at (-1,1,-1){};
		\node[draw, shape=circle, inner sep=4pt] (e) at (0,3.45,0){};
		\node (aa) at (1.5,1,1) {$3$};\node (bb) at (-1.4,-1.4,1) {$2$};\node (cc) at (1.4,-1.4,-1) {$5$};\node (cc) at (-1.4,1.4,-1) {$4$};\node (ee) at (-0.5,3.45,0){$1$};
		\draw (c)--(a)--(d)--(c)--(b);\draw(b)--(d);\draw[white, line width=5pt](a)--(b);\draw(a)--(b);
		\draw (a)--(e)--(d);
		\node (t) at (-2.5,0,0){$\to$};
		\node (s) at (2.5,0,0){$\to$};
	\end{tikzpicture}
\end{minipage}&
\begin{minipage}[b]{0.3\linewidth}
\centering
	\begin{tikzpicture}[scale=0.8]
		\node[draw, shape=circle, inner sep=4pt] (a) at (1,1,1){};
		\node[draw, shape=circle, inner sep=4pt] (b) at (-1,-1,1){};
		\node[draw, shape=circle, inner sep=4pt] (c) at (1,-1,-1){};
		\node[draw, shape=circle, inner sep=4pt] (d) at (-1,1,-1){};
		\node[draw, shape=circle, inner sep=4pt] (e) at (5/3,5/3,-5/3){};
		\node (aa) at (1.5,1,1) {$3$};\node (bb) at (-1.4,-1.4,1) {$2$};\node (cc) at (1.4,-1.4,-1) {$5$};\node (cc) at (-1.4,1.4,-1) {$4$};\node (ee) at (13/6,5/3,-5/3){$1$};
		\draw (c)--(a)--(d)--(c)--(b);\draw(b)--(d);\draw[white, line width=5pt](a)--(b);\draw(a)--(b);
		\draw (a)--(e)--(d);\draw(c)--(e);
	\end{tikzpicture}
\end{minipage}
\end{tabular}
 \caption{Schematic of an isomerization reaction of bi-tetrahedron}
 \label{fig:bitetrahedron}
\end{figure}

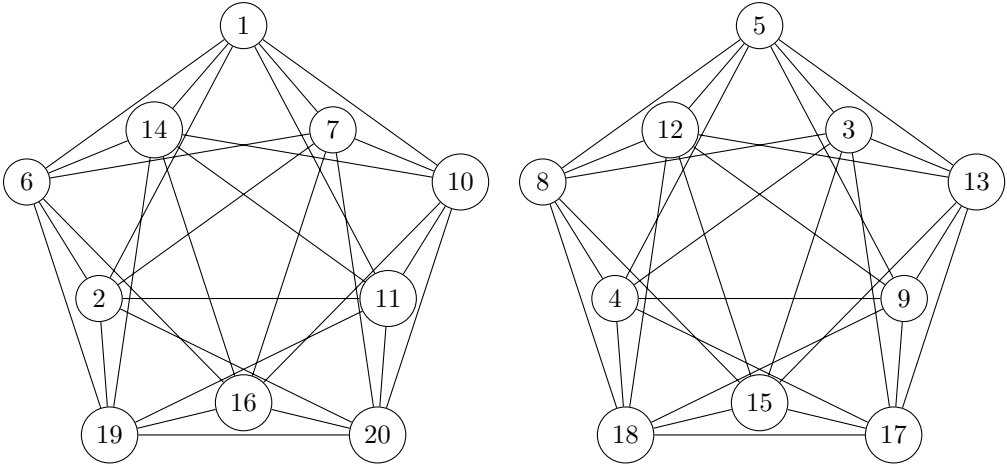
\begin{figure}[htbp]
 \centering

\begin{tabular}{ccc}
\begin{minipage}[b]{0.4\linewidth}\centering\footnotesize
\begin{tikzpicture}[scale=1]
	\node[draw, circle] (a) at ({-3*sin(0)},{3*cos(0)}) {$1$};
	\node[draw, circle] (b) at ({-3*sin(72)},{3*cos(72)}) {$6$};
	\node[draw, circle] (c) at ({-3*sin(144)},{3*cos(144)}) {$19$};
	\node[draw, circle] (d) at ({-3*sin(216)},{3*cos(216)}) {$20$};
	\node[draw, circle] (e) at ({-3*sin(288)},{3*cos(288)}) {$10$};
	\foreach \r in {2}{
	\node[draw, circle] (f) at ({-\r*sin(36)},{\r*cos(36)}) {$14$};
	\node[draw, circle] (g) at ({-\r*sin(108)},{\r*cos(108)}) {$2$};
	\node[draw, circle] (h) at ({-\r*sin(180)},{\r*cos(180)}) {$16$};
	\node[draw, circle] (i) at ({-\r*sin(252)},{\r*cos(252)}) {$11$};
	\node[draw, circle] (j) at ({-\r*sin(324)},{\r*cos(324)}) {$7$};
	}
	\draw (a)--(b)--(c)--(d)--(e)--(a)--(f)--(h)--(j)--(g)--(i)--(f)--(b)--(g)--(c)--(h)--(d)--(i)--(e)--(j)--(a)--(g)--(d)--(j)--(b)--(h)--(e)--(f)--(c)--(i)--(a);
\end{tikzpicture}
\end{minipage}&\begin{minipage}[b]{0.4\linewidth}\centering\footnotesize
\begin{tikzpicture}[scale=1]
	\node[draw, circle] (a) at ({-3*sin(0)},{3*cos(0)}) {$5$};
	\node[draw, circle] (b) at ({-3*sin(72)},{3*cos(72)}) {$8$};
	\node[draw, circle] (c) at ({-3*sin(144)},{3*cos(144)}) {$18$};
	\node[draw, circle] (d) at ({-3*sin(216)},{3*cos(216)}) {$17$};
	\node[draw, circle] (e) at ({-3*sin(288)},{3*cos(288)}) {$13$};
	\foreach \r in {2}{
	\node[draw, circle] (f) at ({-\r*sin(36)},{\r*cos(36)}) {$12$};
	\node[draw, circle] (g) at ({-\r*sin(108)},{\r*cos(108)}) {$4$};
	\node[draw, circle] (h) at ({-\r*sin(180)},{\r*cos(180)}) {$15$};
	\node[draw, circle] (i) at ({-\r*sin(252)},{\r*cos(252)}) {$9$};
	\node[draw, circle] (j) at ({-\r*sin(324)},{\r*cos(324)}) {$3$};
	}
	\draw (a)--(b)--(c)--(d)--(e)--(a)--(f)--(h)--(j)--(g)--(i)--(f)--(b)--(g)--(c)--(h)--(d)--(i)--(e)--(j)--(a)--(g)--(d)--(j)--(b)--(h)--(e)--(f)--(c)--(i)--(a);
\end{tikzpicture}
\end{minipage}
\end{tabular}
 \caption{RRM of the isomerization reaction of the bi-tetrahedron in the shape space. Each vertex corresponds to the $20$ cosets in Eq.~\eqref{eq:lcd} indexed from $1$ to $20$ from the beginning to the end of the equation. For example, the vertex $1$ corresponds to the first coset $\mathfrak{U} \left( r_{\textnormal{EQ}} \right)$. Each edge corresponds to the direct isomerization reaction path.}
 \label{fig:rrm_bitetrahedron}
\end{figure}

Consider an isomerization reaction of bi-tetrahendron in Fig.~\ref{fig:bitetrahedron} consisting of 
\begin{math}
5
\end{math}
identical particles.
From the left to right, we denote the configurations as the reactant, transition state and product, respectively. We denote each configuration in 
\begin{math}
\left( \mathbb{R}^3 \right)^N
\end{math}
as 
\begin{math}
r_R
\end{math}, 
\begin{math}
r_T
\end{math}, and 
\begin{math}
r_P
\end{math}, respectively. Note that they are achiral. We set
\begin{math}
r_{\mathrm{EQ}} = r_R
\end{math}. Then, there exists 
\begin{math}
g \in \mathrm{E}^+ \left( 3 \right)
\end{math}
such that 
\begin{math}
g \cdot r_P = \left( 2, 5 \right) \left( 3, 4 \right) \cdot r_{\mathrm{EQ}} 
\end{math}
holds. In this case, 
\begin{math}
\mathfrak{U} \left( r_{\mathrm{EQ}} \right)
\end{math}
is the subgroup of 
\begin{math}
\mathrm{Sym}_\Omega
\end{math}
generated by 
\begin{math}
\left( 2, 3, 4 \right)
\end{math}
and 
\begin{math}
\left( 1, 5 \right) \left( 3, 4 \right)
\end{math}. The left coset decomposition of 
\begin{math}
\mathrm{Sym}_\Omega
\end{math}
by 
\begin{math}
\mathfrak{U} \left( r_{\mathrm{EQ}} \right)
\end{math}
is 
\begin{align}
\mathrm{Sym}_\Omega = &\mathfrak{U} \left( r_{\mathrm{EQ}} \right) \cup \left( 2,3,5 \right)^{-1} \mathfrak{U} \left( r_{\mathrm{EQ}} \right) \cup \left( 4,5 \right) \mathfrak{U} \left( r_{\mathrm{EQ}} \right) \cup \left( 2,3,4,5 \right)^{-1} \mathfrak{U} \left( r_{\mathrm{EQ}} \right) \nonumber \\
&\cup \left( 3, 4 \right) \mathfrak{U} \left( r_{\mathrm{EQ}} \right) \cup \left( 3, 4, 5 \right)^{-1} \mathfrak{U} \left( r_{\mathrm{EQ}} \right) \cup \left( 3, 5, 4 \right)^{-1} \mathfrak{U} \left( r_{\mathrm{EQ}} \right) \nonumber \\
&\cup \left( 3, 5 \right) \mathfrak{U} \left( r_{\mathrm{EQ}} \right) \cup \left( 1, 2 \right) \mathfrak{U} \left( r_{\mathrm{EQ}} \right) \cup \left( 1, 4, 2 \right)^{-1} \mathfrak{U} \left( r_{\mathrm{EQ}} \right) \cup \left( 1, 2 \right) \left( 3, 4 \right) \mathfrak{U} \left( r_{\mathrm{EQ}} \right) \nonumber \\
&\cup \left( 1, 3, 4, 2 \right)^{-1} \mathfrak{U} \left( r_{\mathrm{EQ}} \right) \cup \left( 1, 4, 3, 2 \right)^{-1} \mathfrak{U} \left( r_{\mathrm{EQ}} \right) \cup \left( 1, 3, 2 \right)^{-1} \mathfrak{U} \left( r_{\mathrm{EQ}} \right) \nonumber \\
&\cup \left( 1, 3, 2 \right)^{-1} \left( 4, 5 \right) \mathfrak{U} \left( r_{\mathrm{EQ}} \right) \cup \left( 1, 3, 5, 4, 2 \right)^{-1} \mathfrak{U} \left( r_{\mathrm{EQ}} \right) \cup \left( 1, 2 \right) \left( 3, 5, 4 \right)^{-1} \mathfrak{U} \left( r_{\mathrm{EQ}} \right) \nonumber \\
&\cup \left( 1, 2 \right) \left( 3, 4, 5 \right)^{-1} \mathfrak{U} \left( r_{\mathrm{EQ}} \right) \cup \left( 1, 2 \right) \left( 3, 5 \right) \mathfrak{U} \left( r_{\mathrm{EQ}} \right) \cup \left( 1, 2 \right) \left( 4, 5 \right) \mathfrak{U} \left( r_{\mathrm{EQ}} \right). \label{eq:lcoset}
\end{align}
There are $20$ cosets and we index the cosets from 
\begin{math}
1
\end{math}
to 
\begin{math}
20
\end{math}
from the beginning to the end of the equation.
\begin{math}
\mathfrak{U} \left( r_T \right)
\end{math}
is the subgroup of 
\begin{math}
\mathrm{Sym}_\Omega
\end{math}
generated by 
\begin{math}
\left( 3, 4 \right) \left( 2, 5 \right)
\end{math}
and the left coset decomposition of 
\begin{math}
\mathrm{Sym}_\Omega
\end{math}
by 
\begin{math}
\mathfrak{U} \left( r_T \right)
\end{math}
comprises
\begin{math}
60
\end{math}
cosets. The RRM in the space can be computed using GAP software (Groups, Algorithms and Programming \cite{GAP4}). For details, see Sec.~2 in Supporting Information. The resulting RRM is shown in Fig.~\ref{fig:rrm_bitetrahedron}. The resulting RRM comprises two connected components. This is because the rotation and reaction shown in Fig.~\ref{fig:bitetrahedron} induce even permutations and thus the parity of the permutation is conserved in the reaction in Fig.~\ref{fig:bitetrahedron}. Based on the results presented in the previous section, the two connected components are isomorphic. Each connected component is the line graph of the complete graph of $5$ vertices. 
\subsection{Isomerization reaction of $\mathrm{PF_5}$ by Berry's pseudo rotation mechanism}
\begin{figure}[htbp]
 \centering
 \includegraphics[keepaspectratio, scale=1.0]{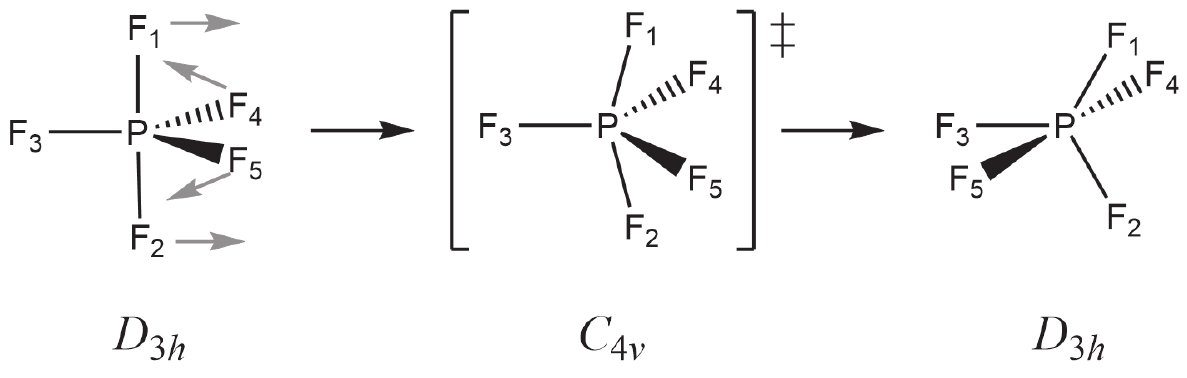}
 \caption{Schematic figure of Berry's pseudo rotation mechanism}
 \label{fig:pseudo}
\end{figure}

Consider the isomerization reaction of $\mathrm{PF_5}$ by Berry's pseudo rotation mechanism in Fig.~\ref{fig:pseudo} \cite{Brocas1983,doi:10.1063/1.1730820}. In this case, it is known that the resulting RRM in the shape space is the Desargues-Levi graph \cite{doi:10.1021/ar50034a001,Pisanski2013}. We demonstrated the proposed algorithm in this system because it is one of the most well-known isomerization. Our results are consistent with those reported in Ref.~\citenum{doi:10.1021/ar50034a001}. 

\begin{figure}[ht]
\centering
  \begin{tikzpicture}[xscale=0.48, yscale=0.5]\scriptsize
%\tikzset{every node/.style={sloped}}
	\node[draw, ellipse] (a) at (-12,6) {$(1,3,2)(4,5)$};
	\node[draw, ellipse] (b) at (-12,3.5) {$(2,3,5)$};
	\node[draw, ellipse] (c) at (-12,1) {$(1,3,4,2)$};
	\node[draw, ellipse] (d) at (-12,-1) {$(1,3)(2,5)$};
	\node[draw, ellipse] (e) at (-12,-3.5) {$(4,5)$};
	\node[draw, ellipse] (f) at (-12,-6) {$(1,3,5,2,4)$};
	\node[draw, ellipse] (g) at (-2,3.5) {$(1,3)(2,5,4)$};
	\node[draw, ellipse] (h) at (-5,-1) {$(2,3,5,4)$};
	\node[draw, ellipse] (i) at (-2,-2.5) {$(1,3,4,5,2)$};
	\node[draw, ellipse] (j) at (-2,-6) {$(2,3)$};
	\node[draw, ellipse] (k) at (2,6) {$(2,3,4)$};
	\node[draw, ellipse] (l) at (2,2.5) {$(1,3,5,2)$};
	\node[draw, ellipse] (m) at (4.6,1) {$(2,3)(4,5)$};
	\node[draw, ellipse] (n) at (2,-3.5) {$(1,3)(2,4)$};
	\node[draw, ellipse] (o) at (12,6) {$(1,3)(2,4,5)$};
	\node[draw, ellipse] (p) at (12,3.5) {$()$};
	\node[draw, ellipse] (q) at (12,1) {$(1,3,4)(2,5)$};
	\node[draw, ellipse] (r) at (12,-1) {$(1,3,2)$};
	\node[draw, ellipse] (s) at (12,-3.5) {$(2,3,4,5)$};
	\node[draw, ellipse] (t) at (12,-6) {$(1,3,5,4,2)$};
	\draw[rounded corners=10pt] (a)--node[auto=right]{$(2,5)(3,4)$}(b)--node[auto=right]{$(2,4,5)$}(c)--node[auto=right]{$(1,3,4,2)$}(d)--node[auto=right]{$(3,4)$}(e)--node[auto=right]{$(4,5)$}(f)--(-16,-6)--(-16,6)--(a);\node[rotate=-90] (x) at (-15.5,0){$(1,3,2)(4,5)$};
	\draw[rounded corners=10pt] (o)--node[auto=left]{$(3,4,5)$}(p)--node[auto=left]{$()$}(q)--node[auto=left]{$(1,3,2)$}(r)--node[auto=left]{$(2,4,3,5)$}(s)--node[auto=left]{$(2,5)$}(t)--(16,-6)--(16,6)--(o);\node[rotate=-90] (y) at (15.5,0){$(1,3,5,4,2)$};
	\draw (a)--node[auto=left]{$(2,4)(3,5)$}(k)--node[auto=left]{$(2,3,4)$}(o);
	\draw (b)--node[auto=left]{$(2,3,5)$}(g)--node[auto=left,pos=0.6]{$(3,4,5)$}(p);
	\draw (c)--node[auto=left, pos=0.25]{$(2,5,3)$}(m)--node[auto=left]{$(2,3)(4,5)$}(q);
	\draw (d)--node[auto=right]{$(2,3,5,4)$}(h)--node[auto=right,, pos=0.75]{$(2,5,3,4)$}(r);
	\draw (e)--node[auto=right, pos=0.4]{$(3,5)$}(n)--node[auto=right]{$(2,3,4,5)$}(s);
	\draw (f)--node[auto=right]{$(2,3)$}(j)--node[auto=right]{$(2,4,5,3)$}(t);
	\draw (g)--node[auto=right, pos=0.1]{$(1,3,4,5,2)$}(i)--node[auto=right, pos=0.8]{$(2,5,4,3)$}(j);
	\draw (k)--node[auto=left, pos=0.2]{$(2,5,4)$}(l)--node[auto=left, pos=0.9]{$(1,3,5,2)$}(n);
	\draw[rounded corners=5pt] (h)--node[auto=right]{$(2,4)$}(-5,-2.5)--(i);
	\draw[rounded corners=5pt] (l)--(4.6,2.5)--node[auto=left]{$(2,4,3)$}(m);
  \end{tikzpicture}
 \caption{RRM of the isomerization reaction of $\mathrm{PF_5}$ by Berry's pseudo rotation mechanism in the shape space. Each vertex corresponds to the permutation isomer of the equilibrium state $\mathrm{PF_5}$ and permutation shown in each vertex corresponds to the permutation from the reference structure (the left most structure in Figure~\ref{fig:pseudo}). Each edge corresponds to the direct isomerization reaction path and its label is the permutation from the reference structure (the middle structure in Figure.~\ref{fig:pseudo}).}
 \label{fig:rrm_pseudo}
\end{figure}
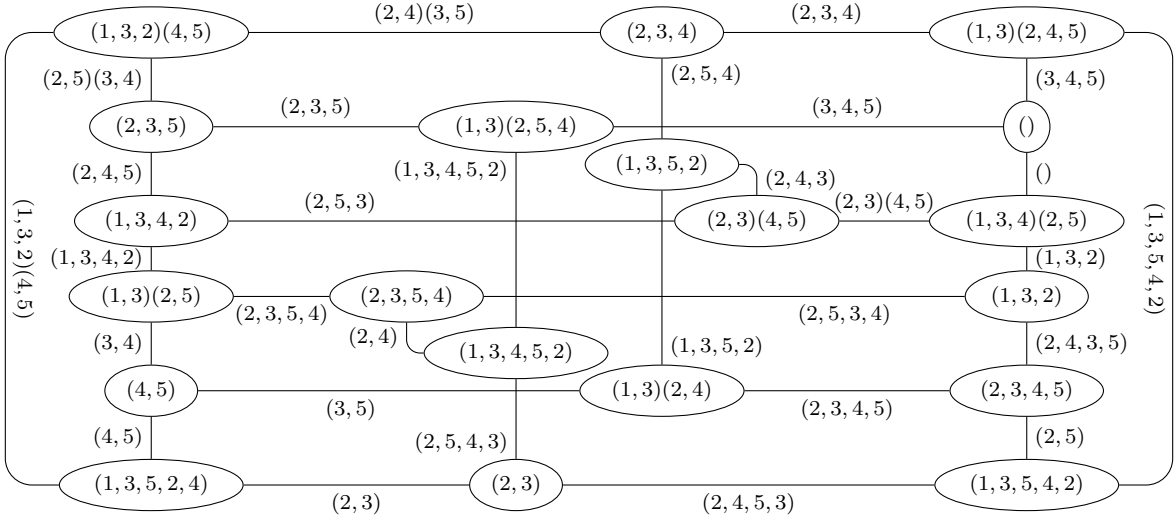

Consider the isomerization of $\mathrm{PF_5}$ by Berry's pseudo rotation mechanism, as shown in Fig.~\ref{fig:pseudo} \cite{Brocas1983,doi:10.1063/1.1730820}.
In this system, we take the left configuration as the reference structure 
\begin{math}
r_{\mathrm{EQ}}
\end{math}
of the reactant and product and the middle configuration 
\begin{math}
r_{\mathrm{T}}
\end{math}
as the reference structure of the transition structure. In this system, 
\begin{math}
\mathrm{Sym}_\Omega \cong \mathfrak{S}_5 \times \mathfrak{S}_1 \cong \mathfrak{S}_5
\end{math}, which is the permutation group acting on the set of the five Fluorine atoms. 
\begin{math}
\mathfrak{U} \left( r_{\mathrm{EQ}} \right)
\end{math}
is the subgroup of 
\begin{math}
\mathrm{Sym}_\Omega
\end{math}
generated by 
\begin{math}
\left( 3, 4, 5 \right)
\end{math}
and 
\begin{math}
\left( 1, 2 \right) \left( 4, 5 \right)
\end{math}. 
\begin{math}
\mathfrak{U} \left( r_{\mathrm{T}} \right)
\end{math}
is the subgroup of 
\begin{math}
\mathrm{Sym}_\Omega
\end{math}
generated by 
\begin{math}
\left( 1, 4, 2, 5 \right)
\end{math}. The right structure (the product structure
\begin{math}
r_P
\end{math}) in Fig.~\ref{fig:pseudo} is related to the reference structure 
\begin{math}
r_{\mathrm{EQ}}
\end{math}
by 
\begin{math}
r_P = \left( 2, 3, 4 \right) \left( 1, 5 \right) \cdot r_{\mathrm{EQ}}
\end{math}. Using them as the input for the Algorithm, we obtain the RRM in Fig.~\ref{fig:rrm_pseudo}.
In the RRM, the labels of the vertices indicate the permutation from the reference structure 
\begin{math}
r_{\mathrm{EQ}}
\end{math}
and those of the edges indicate the permutation from the reference structure 
\begin{math}
r_T
\end{math}. The RRM comprises
\begin{math}
20
\end{math}
vertices and 
\begin{math}
30
\end{math}
edges, which is isomorphic to the Desargues-Levi graph.

\section{Demonstration of Algorithm in reactions of realistic molecules} \label{sec:demo_reals}
In this section, we demonstrate the proposed algorithm in reactions of realistic molecules, taking 
\begin{math}
\mathrm{Au}_5
\end{math}
and Pentane 
\begin{math}
\mathrm{C}_5 \mathrm{H}_{12}
\end{math}
as examples. The former system has equilibrium structures with various types of symmetries; thus, symmetry considerations are important. The latter is one of the most common organic molecules used in chemistry. The order of its 
\begin{math}
\mathrm{Sym}_\Omega
\end{math}
is 
\begin{math}
5! \times 12! = 57480192000
\end{math}, which poses a significant challenge for the efficiency of the proposed algorithm. In this case, it was turned out that the computation of the entire RRM in the shape space is infeasible, but that of a connected component of the RRM is still feasible, demonstrating that the proposed algorithm computing 
\begin{math}
\mathrm{Sym}^{\mathrm{c}}_\Omega \left( \mathrm{G}_0 \right)
\end{math}
is valuable.

In this section, RRMs in 
\begin{math}
\left( \mathbb{R}^3 \right)^N / \mathrm{E} \left( 3 \right) \times \mathrm{Sym}_\Omega
\end{math}
are termed the original RRM and all the inputs are computed by using GRRM program. For details of the input, see \cite{Murayama2022}. In principle, the proposed algorithm can take any inputs computed using the GRRM program, provided that the input does not contain dissociation channels (DCs) and saddle connections, i.e., reaction paths ending up with other saddles, which may occur if the valley-ridge transition \cite{doi:10.1063/1.4935181,doi:10.1063/1.4935182} occurs in the middle of the reaction path. They are also extremely important features of the potential energy landscape and will be considered in the algorithm in our subsequent study. 

Given the output of the GRRM algorithm, \verb|*_EQ_list.log|, \verb|*_TS_list.log|, and  \verb|*_TS*.log|, we extract the list 
\begin{math}
\left\{ \left( r_T^{\left( i \right)}, \left\{ r_R^{\left( i \right)}, r_P^{\left( i \right)} \right\} \right) \right\}_{i \in \tilde{I}}
\end{math}
and compute the RRM in the shape space by using the Algorithm described in Sec.~5. For details on the implementation, see Sec.~3 in Supporting Information.

\subsection{Demonstration in $\mathrm{Au}_5$}
\begin{figure}[htbp]
 \centering
 \includegraphics[keepaspectratio, scale=0.25]
      {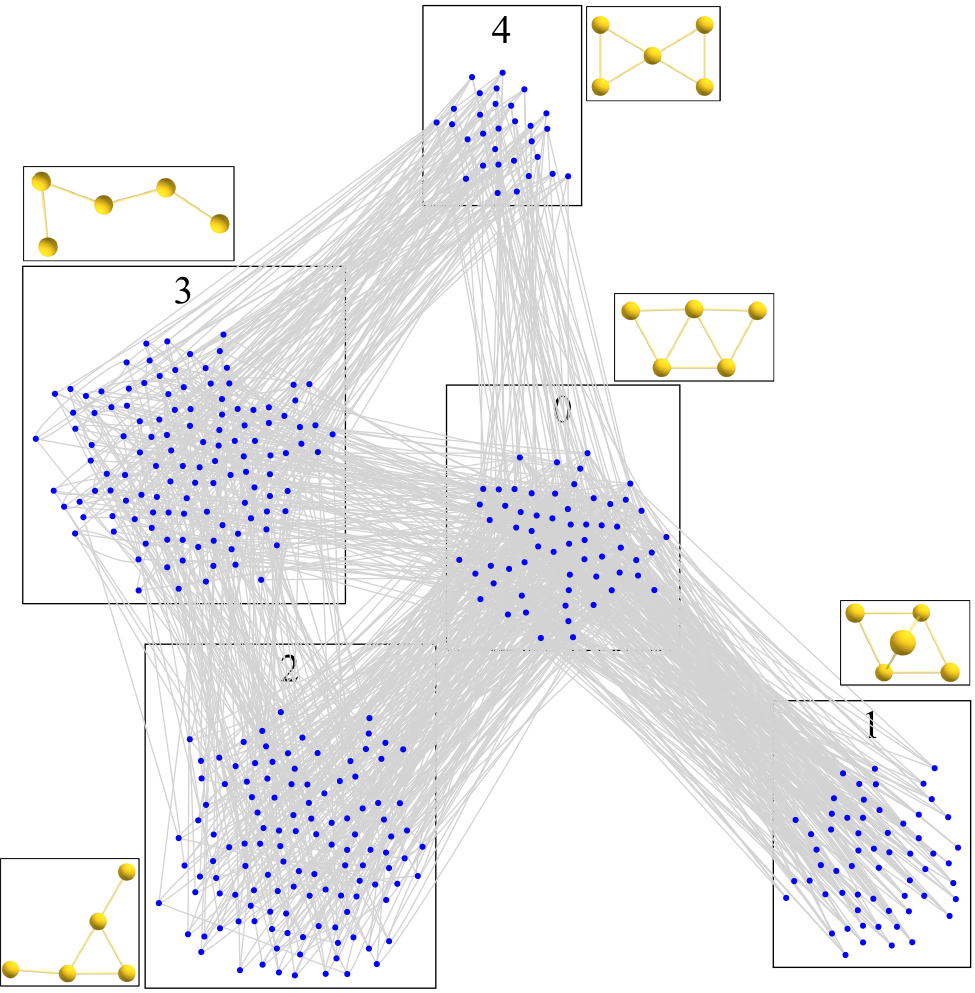}
 \caption{RRM of $\mathrm{Au}_5$ in the shape space. Each blue dot corresponds to a distinct permutation isomer in the shape space and each edge corresponds to a reaction path. Each set of framed vertices corresponds to a structure in the figure numbered from $0$ to $4$.}
 \label{fig:rrm_Au5}
\end{figure}

The original RRM comprises
\begin{math}
5
\end{math}
vertices that are indexed as 
\begin{math}
0, 1, 2, 3, 4
\end{math}. The RRM in the shape space computed by using the proposed algorithm is shown in Fig.~\ref{fig:rrm_Au5}, where the numbers 
\begin{math}
0, 1, 2, 3, 4
\end{math}
correspond to the vertices in the original RRM and the blue points in each box are the permutation isomers of the corresponding equilibrium structure. The number of the blue points in each box is 
\begin{math}
N!
\end{math}
divided by the symmetry number of the corresponding equilibrium structure. The edges inside each box correspond to the self-loops in the original RRM. The number of edges corresponding to each edges in the original RRM is
\begin{math}
N!
\end{math}
divided by the symmetry number of the corresponding transition state structure. 

In this case, all the permutational isomers of 
\begin{math}
\mathrm{Au}_5
\end{math}
are connected by  reaction paths. Recently, Tsutsumi et al studied how permutation-inversion isomers of the conformation $0$ in Fig.~\ref{fig:rrm_Au5} are connected by reaction paths corresponding to transition states of low lying energies and visualize the resulting network in Ref~\citenum{Tsutsumi2022}. They found the second to last energy transition states are enough to obtain an RRM in which all the permutation-inversion isomers of the conformation $0$ are connected. This provides valuable information on the  permutation that is feasible in the sense of Longuet--Higgins \cite{Longuet-Higgins1963}. The proposed algorithm enables automatic construction of such an RRM and makes such a study more systematic.
To quantify the resulting RRM in shape space, we computed the number of cliques and number of independent cycles (the first betti number of RRM). Here a clique in a multi-graph is a subgraph isomorphic to a complete graph. In the inputted RRM has $5$ $1$-cliques, $7$ $2$-cliques, $3$ $3$-cliques and $8$ independent cycles whereas the RRM in the shape space has $390$ $1$-cliques, $1080$ $2$-cliques, $120$ $3$-cliques, and $691$ independent cycles. These characteristics of RRM in the shape space reflect topological features of the potential energy surface in the shape space (roughy that is $(3N-6)$-dimensional coordinate space). These characteristics are useful to compare properties of molecules of various symmetry. For instance, we show these characteristic for 
\begin{math}
\mathrm{Au}_\alpha \mathrm{Cu}_{5-\alpha}
\end{math}
for 
\begin{math}
\alpha \in \left\{ 0, \cdots, 5 \right\}
\end{math}
in Table~\ref{tab:aucu_comparison}. These molecules have different CNPI-groups depending on the compositions and their corresponding symmetry-reduced spaces are different. The characteristic of RRM in shape space offers a fair basis for comparison in such a case. We will announce more detailed analysis in the subsequent paper.
\begin{table}[h]
\centering
\begin{tabular}{|c|c|c|c|c|c|c|}
\hline
& \textbf{1-Clique} & \textbf{2-Clique} & \textbf{3-Clique} & \textbf{4-Clique} & \textbf{5-Clique} & \textbf{Cycle Basis}\\
\hline
$\mathrm{Au}_5$ & 390 & 1080 & 120 & - & - & 691\\
\hline
$\mathrm{Au}_4\mathrm{Cu}$ & 168 & 450 & 78 & - & - & 307\\
\hline
$\mathrm{Au}_3\mathrm{Cu}_2$ & 192 & 597 & 434 & 117 & 12 & 430\\
\hline
$\mathrm{Au}_2\mathrm{Cu}_3$ & 126 & 393 & 186 & 18 & - & 280\\
\hline
$\mathrm{Au}\mathrm{Cu}_4$ & 102 & 330 & 48 & - & - & 253\\
\hline
$\mathrm{Cu}_5$ & 120 & 420 & 420 & 255 & 60 & 301\\
\hline
\end{tabular}
\caption{Number of cliques and cycle basis for different compounds. In the table, the '-' symbol denotes that there is no clique of the size.}
\label{tab:aucu_comparison}
\end{table}

\subsection{Demonstration in $\mathrm{C}_5 \mathrm{H}_{12}$} \label{sec:demoC5H12}
\begin{figure}[htbp]
 \centering
 \includegraphics[keepaspectratio, scale=0.25]
      {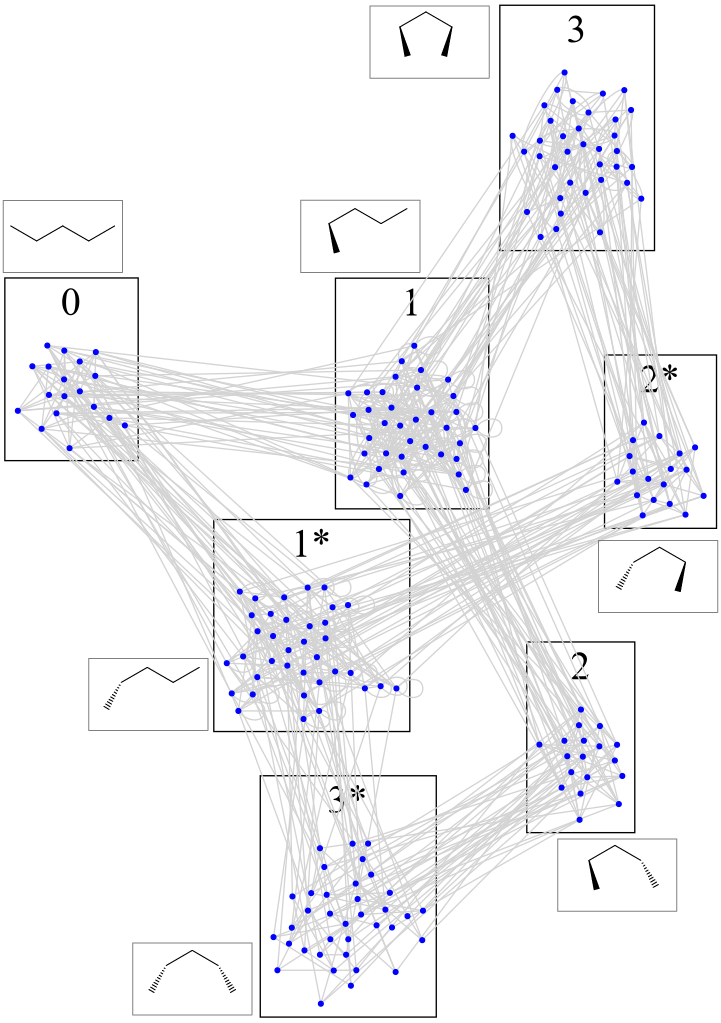}
 \caption{RRM of pentane in the shape space. Each blue dot corresponds to a permutation isomer and each edge corresponds to the reaction path connecting distinct equilibrium structures in the shape space. Each set of framed vertices corresponds to a structure in the figure numbered from $0$ to $3$ where $n*$ is the inversion isomer of $n$ for $n \in \left\{ 1, 2, 3 \right\}$.}
 \label{fig:rrm_pentane}
\end{figure}

The original RRM comprises $3$ vertices indexed as 0, 1, 2, and 3. One of the connected components of the RRM in the shape space is shown in Fig.~\ref{fig:rrm_pentane}, where the numbers correspond to the vertices in the original RRM and $n$* indicates the spatial inversion of the vertex $n$. In this case, the number of connected components of the RRM in the shape space is 
\begin{math}
1596672000
\end{math}, which is equal to 
\begin{math}
\left[ \mathrm{Sym}_\Omega : \mathrm{Sym}_\Omega^c \right]
\end{math}.
All the connected components are mapped to each other by the action of 
\begin{math}
\mathrm{Sym}_\Omega
\end{math}
and thus they are isomorphic in the sense of Def~\ref{def;graph_isomorphism}. Therefore, it is sufficient to investigate a single connected component of the RRM as shown in Fig.~\ref{fig:rrm_pentane}.
\section{Remark on Absolute Rate Theory}
In this section, we derive the rate equation in the symmetry-reduced space 
\begin{math}
\left( \mathbb{R}^3 \right)^N / \mathrm{E}^+ \left( 3 \right) \times \mathrm{Sym}_\Omega
\end{math}
starting from a rate equation in the shape space. Historically, several studies have been conducted on \cite{doi:10.1021/ja00478a010,doi:10.1021/ja00478a009} what should be the correct form of the rate equation in the symmetry-reduced space. Our derivation depends only on the symmetry of the rate equation in the shape space and is not subject to a specific expression of the reaction rate constant. The resulting rate equation in the symmetry-reduced space is consistent with those in Ref.~\citenum{doi:10.1021/ja00478a010,doi:10.1021/ja00478a009}.

Let 
\begin{math}
X_v \left( t \right)
\end{math}
be the probability that the state is in the equilibrium structure 
\begin{math}
v \in \mathrm{V}
\end{math}
at the time 
\begin{math}
t \in \mathbb{R}
\end{math}. We suppose 
\begin{math}
X = \left\{ X_v \right\}
\end{math}
obeys the following ordinary differential equation, known as the absolute rate equation.
\begin{equation}
\frac{dX_v \left( t \right)}{dt} = \sum_{v' \in \mathrm{V}} k_{v,v'} X_{v'} \left( t \right),
\end{equation}
for 
\begin{math}
v \in \mathrm{V}
\end{math}. 
We suppose
\begin{math}
k_{v,v'}
\end{math}
is a real number that satisfies 
\begin{math}
k_{\sigma \cdot v, \sigma \cdot v'} = k_{v,v'}
\end{math}
holds for all 
\begin{math}
v, v' \in \mathrm{V}
\end{math}
and 
\begin{math}
\sigma \in \mathrm{Sym}_\Omega
\end{math}. Let 
\begin{math}
Y_u \left( t \right) = \sum_{v \in \mathrm{V}, \pi_{\mathrm{V}} \left( v \right) = u} X_v \left( t \right)
\end{math}
be the sum of the probabilities of the states 
\begin{math}
v \in \mathrm{V}
\end{math}
whose projection is 
\begin{math}
u \in \pi_\mathrm{V} \left( \mathrm{V} \right)
\end{math}, i.e., the probability that the state is in 
\begin{math}
u
\end{math}
in the symmetry-reduced space.
\begin{math}
Y_u \left( t \right)
\end{math}
satisfies the following equation. First, note that  
\begin{align}
\frac{d Y_u \left( t \right)}{dt} &= \sum_{v \in \mathrm{V}, \pi_{\mathrm{V}} \left( v \right) = u} \frac{d X_v \left( t \right)}{dt}, \\
&= \sum_{v \in \mathrm{V}, \pi_{\mathrm{V}} \left( v \right) = u} \sum_{v' \in \mathrm{V}} k_{v,v'} X_{v'} \left( t \right) = \sum_{v' \in \mathrm{V}} \left( \sum_{v \in \mathrm{V}, \pi_{\mathrm{V}} \left( v \right) = u} k_{v,v'} \right) X_{v'} \left( t \right) \label{eq:dYdt}
\end{align}
holds. Since 
\begin{equation}
\sum_{v \in \mathrm{V}, \pi_{\mathrm{V}} \left( v \right) = u} k_{v, \sigma \cdot v'} = \sum_{v \in \mathrm{V}, \pi_{\mathrm{V}} \left( v \right) = u} k_{\sigma^{-1} \cdot v, v'} = \sum_{v \in \mathrm{V}, \pi_{\mathrm{V}} \left( v \right) = u} k_{v, v'}
\end{equation}
holds for all 
\begin{math}
u \in \pi_{\mathrm{V}} \left( \mathrm{V} \right)
\end{math}, 
\begin{math}
v' \in \mathrm{V}
\end{math}
and 
\begin{math}
\sigma \in \mathrm{Sym}_\Omega
\end{math}, Eq.~\eqref{eq:dYdt} can be written as 
\begin{align}
\sum_{v' \in \mathrm{V}} \left( \sum_{\substack{v \in \mathrm{V}, \\ \pi_{\mathrm{V}} \left( v \right) = u}} k_{v,v'} \right) X_{v'} \left( t \right) &= \sum_{u' \in \pi_{\mathrm{V}} \mathrm{V}} \sum_{\left[ \sigma \right] \in \mathrm{Sym}_\Omega / \mathfrak{U} \left( c \left( u' \right) \right)} \left( \sum_{\substack{v \in \mathrm{V}, \\ \pi_{\mathrm{V}} \left( v \right) = u}} k_{v,\sigma \cdot \left[ c \left( u' \right) \right]} \right) X_{\sigma \cdot \left[ c \left( u' \right) \right]} \left( t \right), \\
&= \sum_{u' \in \pi_{\mathrm{V}} \mathrm{V}} \sum_{\left[ \sigma \right] \in \mathrm{Sym}_\Omega / \mathfrak{U} \left( c \left( u' \right) \right)} \left( \sum_{\substack{v \in \mathrm{V}, \\ \pi_{\mathrm{V}} \left( v \right) = u}} k_{v, \left[ c \left( u' \right) \right]} \right) X_{\sigma \cdot \left[ c \left( u' \right) \right]} \left( t \right), \\
&= \sum_{u' \in \pi_{\mathrm{V}} \mathrm{V}} \left( \sum_{\substack{v \in \mathrm{V}, \\ \pi_{\mathrm{V}} \left( v \right) = u}} k_{v,\left[ c \left( u' \right) \right]} \right) \sum_{\left[ \sigma \right] \in \mathrm{Sym}_\Omega / \mathfrak{U} \left( c \left( u' \right) \right)} X_{\sigma \cdot \left[ c \left( u' \right) \right]} \left( t \right), \\
&= \sum_{u' \in \pi_{\mathrm{V}} \mathrm{V}} \left( \sum_{\substack{v \in \mathrm{V}, \\ \pi_{\mathrm{V}} \left( v \right) = u}} k_{v,\left[ c \left( u' \right) \right]} \right) \sum_{v' \in \mathrm{V}, \pi_{\mathrm{V}} \left( v' \right) = u'} X_{v'} \left( t \right), \\
&= \sum_{u' \in \pi_{\mathrm{V}} \mathrm{V}} \left( \sum_{\substack{v \in \mathrm{V}, \\ \pi_{\mathrm{V}} \left( v \right) = u}} k_{v,\left[ c \left( u' \right) \right]} \right) Y_{u'} \left( t \right).
\end{align}
In summary, we obtain
\begin{equation}
\frac{d Y_u \left( t \right)}{dt} = \sum_{u' \in \pi_{\mathrm{V}} \mathrm{V}} \left( \sum_{\substack{v \in \mathrm{V}, \\ \pi_{\mathrm{V}} \left( v \right) = u}} k_{v,\left[ c \left( u' \right) \right]} \right) Y_{u'} \left( t \right). \label{eq:lrate}
\end{equation}
If we define 
\begin{equation}
k_{v,v'} = \sum_{\substack{e \in \mathrm{E}, \\ h \left( e \right) = \left\{ v, v' \right\}}} k_{v,v'}^e
\end{equation}
with
\begin{math}
k^e_{v, v'}
\end{math}
that satisfies
\begin{math}
k^{\sigma \cdot e}_{\sigma \cdot v, \sigma \cdot v'} = k^e_{v, v'}
\end{math}
for
\begin{math}
e \in \mathrm{E}
\end{math}
and 
\begin{math}
v, v' \in \mathrm{V}
\end{math}, 
\begin{equation}
k_{\sigma \cdot v, \sigma \cdot v'} = \sum_{\substack{e \in \mathrm{E}, \\ h \left( e \right) = \left\{ \sigma \cdot v, \sigma \cdot v' \right\}}} k_{\sigma \cdot v, \sigma \cdot v'}^e = \sum_{\substack{e \in \mathrm{E}, \\ h \left( \sigma^{-1} \cdot e \right) = \left\{ v, v' \right\}}} k_{v, v'}^{\sigma^{-1} \cdot e} = \sum_{\substack{e \in \sigma \cdot \mathrm{E}, \\ h \left( e \right) = \left\{ v, v' \right\}}} k_{v, v'}^e = \sum_{\substack{e \in \mathrm{E}, \\ h \left( e \right) = \left\{ v, v' \right\}}} k_{v, v'}^e
\end{equation}
holds for all 
\begin{math}
\sigma \in \mathrm{Sym}_\Omega
\end{math}
using the fact that 
\begin{math}
h
\end{math}
is 
\begin{math}
\mathrm{Sym}_\Omega
\end{math}-equivariant and 
\begin{math}
\sigma \cdot \mathrm{E} = \mathrm{E}
\end{math}. In this case, the absolute rate constant from the state 
\begin{math}
u' \in \pi_{\mathrm{V}} \mathrm{V}
\end{math}
to 
\begin{math}
u \in \pi_{\mathrm{V}} \mathrm{V}
\end{math}
is 
\begin{equation}
\sum_{\substack{v \in \mathrm{V}, \\ \pi_{\mathrm{V}} \left( v \right) = u}} \left( \sum_{\substack{e \in \mathrm{E}, \\ h \left( e \right) = \left\{ v, \left[ c \left( u' \right) \right] \right\}}} k_{v,\left[ c \left( u' \right) \right]}^e \right)
\end{equation}
using Eq.~\eqref{eq:lrate}. This satisfies the following equation.
\begin{align}
\sum_{\substack{v \in \mathrm{V}, \\ \pi_{\mathrm{V}} \left( v \right) = u}} \sum_{\substack{e \in \mathrm{E}, \\ h \left( e \right) = \left\{ v, \left[ c \left( u' \right) \right] \right\}}} k_{v,\left[ c \left( u' \right) \right]}^e &=  \frac{\sigma_u}{\left| \mathrm{Sym}_\Omega \right|}\sum_{\substack{v' \in \mathrm{V}, \\ \pi_{\mathrm{V}} \left( v' \right) = u'}} \sum_{\substack{v \in \mathrm{V}, \\ \pi_{\mathrm{V}} \left( v \right) = u}} \sum_{\substack{e \in \mathrm{E}, \\ h \left( e \right) = \left\{ v, v' \right\}}} k_{v,v'}^e, \\
&=  \frac{\sigma_{u'}}{\left| \mathrm{Sym}_\Omega \right|}\sum_{\substack{v' \in \mathrm{V}, \\ \pi_{\mathrm{V}} \left( v' \right) = u'}} \sum_{\substack{v \in \mathrm{V}, \\ \pi_{\mathrm{V}} \left( v \right) = u}} \sum_{\substack{\check{e} \in \pi_{\mathrm{E}}\mathrm{E}, \\ h_\pi \left( \check{e} \right) = \left\{ u, u' \right\}}} \sum_{\substack{\left[ \sigma \right] \in \mathrm{Sym}_\Omega / \mathfrak{U} \left( c \left( \check{e} \right) \right), \\ h \left( \sigma \cdot c \left( \check{e} \right) \right) = \left\{ v, v' \right\}}} k_{v,v'}^{\sigma \cdot \left[ c \left( \check{e} \right) \right]}, \\
&=  \sum_{\substack{\check{e} \in \pi_{\mathrm{E}}\mathrm{E}, \\ h_\pi \left( \check{e} \right) = \left\{ u, u' \right\}}} \frac{\sigma_{u'}}{\left| \mathrm{Sym}_\Omega \right|}\sum_{\substack{v' \in \mathrm{V}, \\ \pi_{\mathrm{V}} \left( v' \right) = u'}} \sum_{\substack{v \in \mathrm{V}, \\ \pi_{\mathrm{V}} \left( v \right) = u}}\sum_{\substack{\left[ \sigma \right] \in \mathrm{Sym}_\Omega / \mathfrak{U} \left( c \left( \check{e} \right) \right), \\ h \left( \sigma \cdot c \left( \check{e} \right) \right) = \left\{ v, v' \right\}}} k_{v,v'}^{\sigma \cdot \left[ c \left( \check{e} \right) \right]}, \\
&=  \sum_{\substack{\check{e} \in \pi_{\mathrm{E}}\mathrm{E}, \\ h_\pi \left( \check{e} \right) = \left\{ u, u' \right\}}} \frac{\sigma_{u'}}{\left| \mathrm{Sym}_\Omega \right|}\sum_{\substack{v' \in \mathrm{V}, \\ \pi_{\mathrm{V}} \left( v' \right) = u'}} \sum_{\substack{v \in \mathrm{V}, \\ \pi_{\mathrm{V}} \left( v \right) = u}}\sum_{\substack{\left[ \sigma \right] \in \mathrm{Sym}_\Omega / \mathfrak{U} \left( c \left( \check{e} \right) \right), \\ h \left( c \left( \check{e} \right) \right) = \left\{ \sigma^{-1} \cdot v, \sigma^{-1} \cdot v' \right\}}} k_{\sigma^{-1} \cdot v,\sigma^{-1} \cdot v'}^{\left[ c \left( \check{e} \right) \right]}, \\
&=  \sum_{\substack{\check{e} \in \pi_{\mathrm{E}}\mathrm{E}, \\ h_\pi \left( \check{e} \right) = \left\{ u, u' \right\}}} \frac{\sigma_{u'}}{\left| \mathrm{Sym}_\Omega \right|}\sum_{\substack{v' \in \sigma \cdot \mathrm{V}, \\ \pi_{\mathrm{V}} \left( v' \right) = u'}} \sum_{\substack{v \in \sigma \cdot \mathrm{V}, \\ \pi_{\mathrm{V}} \left( v \right) = u}}\sum_{\substack{\left[ \sigma \right] \in \mathrm{Sym}_\Omega / \mathfrak{U} \left( c \left( \check{e} \right) \right), \\ h \left( c \left( \check{e} \right) \right) = \left\{ v, v' \right\}}} k_{v,v'}^{\left[ c \left( \check{e} \right) \right]}, \\
&=  \sum_{\substack{\check{e} \in \pi_{\mathrm{E}}\mathrm{E}, \\ h_\pi \left( \check{e} \right) = \left\{ u, u' \right\}}} \frac{\sigma_{u'}}{\left| \mathrm{Sym}_\Omega \right|}\sum_{\substack{v' \in \mathrm{V}, \\ \pi_{\mathrm{V}} \left( v' \right) = u'}} \sum_{\substack{v \in \mathrm{V}, \\ \pi_{\mathrm{V}} \left( v \right) = u}}\sum_{\substack{\left[ \sigma \right] \in \mathrm{Sym}_\Omega / \mathfrak{U} \left( c \left( \check{e} \right) \right), \\ h \left( c \left( \check{e} \right) \right) = \left\{ v, v' \right\}}} k_{v,v'}^{\left[ c \left( \check{e} \right) \right]}, \\
&=  \sum_{\substack{\check{e} \in \pi_{\mathrm{E}}\mathrm{E}, \\ h_\pi \left( \check{e} \right) = \left\{ u, u' \right\}}} \frac{\sigma_{u'}}{\left| \mathrm{Sym}_\Omega \right|} \frac{\left| \mathrm{Sym}_\Omega \right|}{\sigma_{\check{e}}} \sum_{\substack{v, v' \in \mathrm{V}, \\ \pi_{\mathrm{V}} \left( v \right) = u, \pi_{\mathrm{V}} \left( v' \right) = u', \\ h \left( \left[ c \left( \check{e} \right) \right] \right) = \left\{ v, v' \right\}}} k_{v,v'}^{\left[ c \left( \check{e} \right) \right]}, \\
&=  \sum_{\substack{\check{e} \in \pi_{\mathrm{E}}\mathrm{E}, \\ h_\pi \left( \check{e} \right) = \left\{ u, u' \right\}}} \frac{\sigma_{u'}}{\sigma_{\check{e}}} \sum_{\substack{v, v' \in \mathrm{V}, \\ \pi_{\mathrm{V}} \left( v \right) = u, \pi_{\mathrm{V}} \left( v' \right) = u', \\ h \left( \left[ c \left( \check{e} \right) \right] \right) = \left\{ v, v' \right\}}} k_{v,v'}^{\left[ c \left( \check{e} \right) \right]}, \\
&=  \sum_{\substack{\check{e} \in \pi_{\mathrm{E}}\mathrm{E}, \\ h_\pi \left( \check{e} \right) = \left\{ u, u' \right\}}} \frac{\sigma_{u'}}{\sigma_{\check{e}}} k_{h \left( \left[ c \left( \check{e} \right) \right] \right)_{u},h \left( \left[ c \left( \check{e} \right) \right] \right)_{u'}}^{\left[ c \left( \check{e} \right) \right]}, 
\end{align}
where 
\begin{math}
h \left( \left[ c \left( \check{e} \right) \right] \right)_{u}
\end{math}
is an element of 
\begin{math}
h \left( \left[ c \left( \check{e} \right) \right] \right)
\end{math}
whose projection is 
\begin{math}
u
\end{math}. This implies that the reaction rate constant of the reaction starting from 
\begin{math}
u'
\end{math}
to 
\begin{math}
u
\end{math}
through the edge 
\begin{math}
\check{e}
\end{math}
is the reaction rate constant of one of the representing reactions in the shape space 
\begin{math}
k_{h \left( \left[ c \left( \check{e} \right) \right] \right)_{u},h \left( \left[ c \left( \check{e} \right) \right] \right)_{u'}}^{\left[ c \left( \check{e} \right) \right]}
\end{math}
multiplied by the ratio of the symmetry number of the reactant 
\begin{math}
u'
\end{math}
and that of the transition state 
\begin{math}
\check{e}
\end{math}. This is consistent with the results in Ref.~\citenum{doi:10.1021/ja00478a010,doi:10.1021/ja00478a009}. Note that our derivation is purely based on the symmetry of the rate equation in the shape space and is not subject to any specific expression of the reaction rate constants, which clarify the origin of the correction factor coming from the symmetry numbers.

\section{Conclusion and Future Perspectives}
This study developed an algorithm to reproduce RRMs in the shape space from the outputs of potential search algorithms. The proposed algorithm does not require any encoding of the molecular configurations and is thus applicable to complicated realistic molecules for which efficient encoding is not readily available. To demonstrate this, the GRRM is utilized; however, in principle, it should work with other potential search algorithms. We have shown subgraphs of RRM mapped to each other by the action of the symmetry group are isomorphic and also provided an algorithm to compute the set of feasible permutations. The proposed algorithm was demonstrated in toy models and in more realistic molecules. Moreover, the absolute rate theory was discussed from our perspective. In principle, our implementation can take any input computed using the GRRM program, provided that the input does not contain DCs and saddle connections, i.e., reaction paths ending up with other saddles, which may occur if the valley-ridge transition \cite{doi:10.1063/1.4935181,doi:10.1063/1.4935182} occurs in the middle of the reaction path. These are extremely important features of the potential energy landscape and will be considered in the algorithm in our subsequent study. 

\section{Acknowledgement}
This work was supported in part by the Institute for Quantum Chemical Exploration (IQCE), JSPS KAKENHI for Transformative Research Areas "Hyper-ordered Structures Science" (Grant Number: JP21H05544 and JP23H04093 to M.K.), for Scientifc Research (Grant Number: JP23H01915, JP23KJ0031), the Photo-excitonix Project of Hokkaido University, and JST CREST Grant Number JPMJCR18K3, Japan. Some of the reported calculations were performed using computer facilities at the Research Center for Computational Science, Okazaki (Projects: 21-IMS-C018 and 22-IMS-C019), and at the Research Institute for Information Technology, Kyushu University, Japan. 
The Institute for Chemical Reaction Design and Discovery (ICReDD) was established by the World Premier International Research Initiative (WPI) of MEXT, Japan.

\begin{suppinfo}

The Supporting Information is available free of charge via the Internet at http://pubs.acs.org.

\begin{itemize}
  \item An algorithm to compute a single connected component of the entire graph $G$
  \item GAP program to reproduce the RRM of the isomerization in Fig.~\ref{fig:bitetrahedron}
  \item Implementation detail of the computation of RRM in the shape space from outputs of GRRM program
\end{itemize}

\end{suppinfo}

\bibliography{dyn_recomb}

\end{document}

\maketitle

\section{An algorithm to compute a single connected component of the entire graph $G$} \label{sec:connectedalgorithm}

In this section, we use the terminologies in Sec.~4 in the main text.
Let 
\begin{math}
\mathrm{G}_0 = \left( \mathrm{V}_0, \mathrm{E}_0, h, \pi_\mathrm{V}, \pi_\mathrm{E} \right)
\end{math}
be a connected component of 
\begin{math}
\mathrm{G}
\end{math}.
Let
\begin{math}
\pi \mathrm{G}_0 = \left( \pi_{\mathrm{V}} \mathrm{V}_0, \pi_{\mathrm{E}} \mathrm{E}_0, h_{\pi}, \mathrm{id}_{\mathrm{V}}, \mathrm{id}_{\mathrm{E}} \right)
\end{math}
be the projected graph of 
\begin{math}
\mathrm{G}_0
\end{math}
where 
\begin{math}
h_{\pi} \colon \pi_{\mathrm{E}} \mathrm{E}_0 \rightarrow 2^{\pi_{\mathrm{V}} \mathrm{V}_0}
\end{math}
is a map induced by the 
\begin{math}
\mathrm{Sym}_\Omega
\end{math}-invariant map
\begin{math}
\pi_{\mathrm{V}} \circ h \colon \mathrm{E}_0 \rightarrow 2^{\pi_{\mathrm{V}} \mathrm{V}_0}
\end{math}. In the following, we assume 
\begin{math}
\pi \mathrm{G}
\end{math}
is connected. Otherwise consider each connected component separately. In this case, all the connected components of 
\begin{math}
\mathrm{G}
\end{math}
are in the single 
\begin{math}
\mathrm{Sym}_\Omega
\end{math}-orbit in the space of subgraphs of 
\begin{math}
\mathrm{G}
\end{math}.

Let 
\begin{math}
\mathrm{Sym}^{\mathrm{c}}_{\Omega} \left( \mathrm{G}_0 \right) = \left\{ \sigma \in \mathrm{Sym}_\Omega \middle| \sigma \cdot \mathrm{G}_0 = \mathrm{G}_0 \right\}
\end{math}
be the stabilizer of 
\begin{math}
\mathrm{G}_0
\end{math}. Then, 
\begin{math}
\mathrm{G}
\end{math}
has 
\begin{math}
\left[ \mathrm{Sym}_\Omega : \mathrm{Sym}^{\mathrm{c}}_{\Omega} \left( \mathrm{G}_0 \right) \right]
\end{math}
connected components, all of which are isomorphic. If 
\begin{math}
\mathrm{Sym}^{\mathrm{c}}_{\Omega} \left( \mathrm{G}_0 \right)
\end{math}
is used instead of 
\begin{math}
\mathrm{Sym}_\Omega
\end{math}
in the algorithm described in the previous section, we obtain a graph isomorphic to 
\begin{math}
\mathrm{G}_0
\end{math}. Therefore, the remainder of this section we propose an algorithm to compute 
\begin{math}
\mathrm{Sym}^{\mathrm{c}}_{\Omega} \left( \mathrm{G}_0 \right)
\end{math}. Let 
\begin{math}
v_0
\end{math}
be a vertex in 
\begin{math}
\mathrm{G}_0
\end{math}
and 
\begin{equation}
\mathrm{Sym}^{\mathrm{c}}_\Omega \left( v_0 \right) = \left\{ \sigma \in \mathrm{Sym}_\Omega \middle| v_0 \; \textnormal{and} \; \sigma \cdot v_0 \; \textnormal{in the same connected component.} \right\}.
\end{equation}
If 
\begin{math}
v_1
\end{math}
is another vertex in 
\begin{math}
\mathrm{G}_0
\end{math}, then, 
\begin{math}
\mathrm{Sym}^{\mathrm{c}}_\Omega \left( v_1 \right) = \mathrm{Sym}^{\mathrm{c}}_\Omega \left( v_0 \right)
\end{math}
holds. This can be shown as follows. Suppose 
\begin{math}
v_0
\end{math}
and 
\begin{math}
v_1
\end{math}
are in the same connected component. Take an arbitrary 
\begin{math}
\sigma \in \mathrm{Sym}^{\mathrm{c}}_\Omega \left( v_0 \right)
\end{math}. Then, 
\begin{math}
\sigma \cdot v_0
\end{math}
and 
\begin{math}
\sigma \cdot v_1
\end{math}
are in the same connected component 
\begin{math}
\sigma \cdot \mathrm{G}_0
\end{math}. Since 
\begin{math}
v_0
\end{math}
and
\begin{math}
\sigma \cdot v_0
\end{math}
are in the same connected component by the definition and 
\begin{math}
v_0
\end{math}
and 
\begin{math}
v_1
\end{math}
are in the same connected component by the assumption, 
\begin{math}
v_1
\end{math}
and 
\begin{math}
\sigma \cdot v_1
\end{math}
are in the same connected component. This proves 
\begin{math}
\sigma \in \mathrm{Sym}^{\mathrm{c}}_\Omega \left( v_1 \right)
\end{math}. By swapping the role of 
\begin{math}
v_0
\end{math}
and 
\begin{math}
v_1
\end{math}, we obtain the assertion. By the definition of the stabilizer, we obtain
\begin{math}
\mathrm{Sym}^{\mathrm{c}}_\Omega \left( v_0 \right) = \mathrm{Sym}^{\mathrm{c}}_\Omega \left( \mathrm{G}_0 \right)
\end{math}
and the computational problem is reduced to the computation of 
\begin{math}
\mathrm{Sym}^{\mathrm{c}}_\Omega \left( v_0 \right)
\end{math}.

Let 
\begin{math}
\tilde{\mathrm{G}}_0
\end{math}
be a graph with the vertex sets 
\begin{math}
\tilde{\mathrm{V}}_0 = \pi_\mathrm{V} \mathrm{V}_0 \times \mathrm{Sym}^{\mathrm{c}}_\Omega \left( \mathrm{G}_0 \right)
\end{math}
and the edge sets 
\begin{math}
\tilde{\mathrm{E}}_0 = \pi_\mathrm{E} \mathrm{E}_0 \times \mathrm{Sym}^{\mathrm{c}}_\Omega \left( \mathrm{G}_0 \right)
\end{math}. We define 
\begin{math}
\tilde{h} \colon \tilde{\mathrm{E}}_0 \rightarrow 2^{\tilde{\mathrm{V}}_0}
\end{math}
as follows: Let 
\begin{math}
c \colon \left( \mathbb{R}^3 \right)^N / \mathrm{E}^+ \left( 3 \right) \times \mathrm{Sym}_\Omega \rightarrow \left( \mathbb{R}^3 \right)^N
\end{math}
be a mapping that corresponds each class in 
\begin{math}
\left( \mathbb{R}^3 \right)^N / \mathrm{E}^+ \left( 3 \right) \times \mathrm{Sym}_\Omega
\end{math}
to one of its representative in 
\begin{math}
\left( \mathbb{R}^3 \right)^N
\end{math}. 
For each edge 
\begin{math}
e \in \pi_\mathrm{E} \mathrm{E}_0
\end{math}, set
\begin{math}
\left\{ u_R, u_P \right\} = h_\pi \left( e \right)
\end{math}
for 
\begin{math}
u_R, u_P \in \pi_\mathrm{V} \mathrm{V}_0
\end{math}. Fix 
\begin{math}
\sigma_R, \sigma_P \in \mathrm{Sym}^{\mathrm{c}}_\Omega \left( \mathrm{G}_0 \right)
\end{math}
such that 
\begin{math}
h \left( \left[ c \left( e \right) \right] \right) = \left\{ \left[ \sigma_R \cdot c \left( u_R \right) \right], \left[ \sigma_P \cdot c \left( u_P \right) \right] \right\}
\end{math}
holds. Then, for each 
\begin{math}
\sigma \in \mathrm{Sym}^{\mathrm{c}}_\Omega \left( \mathrm{G}_0 \right)
\end{math}, define 
\begin{math}
\tilde{h} \left( e, \sigma \right) = \left\{ \left( u_R, \sigma \cdot \sigma_R \right), \left( u_P, \sigma \cdot \sigma_P \right) \right\}
\end{math}.
Then, 
\begin{math}
\tilde{\mathrm{G}}_0 = \left( \tilde{\mathrm{V}}, \tilde{\mathrm{E}}, \tilde{h} \right)
\end{math}
becomes a multi-graph. 

Up to this point, we obtain the following sequence of projections:
\begin{equation}
\tilde{\mathrm{G}}_0 \xrightarrow{\pi'} \mathrm{G}_0 \xrightarrow{\pi} \pi \mathrm{G}_0
\end{equation}
where 
\begin{math}
\pi'
\end{math}
is defined as 
\begin{math}
\pi' \left( u, \sigma \right) = \left[ \sigma \cdot c \left( u \right) \right]
\end{math}
and 
\begin{math}
\pi' \left( e, \sigma \right) = \left[ \sigma \cdot c \left( e \right) \right]
\end{math}
for 
\begin{math}
\left( u, \sigma \right) \in \tilde{V}_0
\end{math}
and 
\begin{math}
\left( e, \sigma \right) \in \tilde{E}_0
\end{math}.

By fixing an orientation of the edges of 
\begin{math}
\pi \mathrm{G}_0
\end{math}
and then introducing the orientation of the edges of 
\begin{math}
\tilde{\mathrm{G}}_0
\end{math}
so that the projection 
\begin{math}
\pi \circ \pi'
\end{math}
preserves the orientation, we can regard 
\begin{math}
\tilde{\mathrm{G}}_0
\end{math}
and 
\begin{math}
\pi \mathrm{G}_0
\end{math}
as $1$-dimensional CW complex as described in Section 1.A in Ref.~\citenum{Hatcher} and the projection 
\begin{math}
\pi \circ \pi' \colon \tilde{\mathrm{G}} \rightarrow \pi \mathrm{G}_0
\end{math}
is a fiber bundle with the fiber 
\begin{math}
\mathrm{Sym}^{\mathrm{c}}_\Omega \left( \mathrm{G}_0 \right)
\end{math}, since 
\begin{math}
\mathrm{Sym}^{\mathrm{c}}_\Omega \left( \mathrm{G}_0 \right)
\end{math}
acts freely to 
\begin{math}
\tilde{\mathrm{G}}_0
\end{math}
and 
\begin{math}
\tilde{\mathrm{G}}_0 / \mathrm{Sym}^{\mathrm{c}}_\Omega \left( \mathrm{G}_0 \right)
\end{math}
equals to 
\begin{math}
\pi \mathrm{G}_0
\end{math}. Take an arbitrary vertex 
\begin{math}
\tilde{v}_0
\end{math}
in 
\begin{math}
\tilde{\mathrm{G}}_0
\end{math}. Let 
\begin{math}
C
\end{math}
be a cycle in 
\begin{math}
\pi \mathrm{G}_0
\end{math}
with the endpoint 
\begin{math}
\pi \circ \pi' \left( \tilde{v}_0 \right)
\end{math}. Then, there exists a unique lifted path 
\begin{math}
\tilde{C}
\end{math}
in 
\begin{math}
\tilde{\mathrm{G}}
\end{math}
such that 
\begin{math}
\pi \circ \pi' \tilde{C} = C
\end{math}
and 
\begin{math}
\tilde{C}
\end{math}
starts from 
\begin{math}
\tilde{v}_0
\end{math}. Let 
\begin{math}
\tilde{v}_1
\end{math}
be the endpoint of 
\begin{math}
\tilde{C}
\end{math}. By using Proposition~4.48 in Ref.~\citenum{Hatcher}, the fiber bundle has the homotopy lifting property and thus 
\begin{math}
\tilde{v}_1
\end{math}
depends only on the homotopy class of 
\begin{math}
C
\end{math}, 
\begin{math}
\left[ C \right]
\end{math}.

Let 
\begin{math}
\tilde{v}_0 = \left( u_0, \sigma_0 \right)
\end{math}
and 
\begin{math}
C_1, C_2, \cdots, C_{N_c}
\end{math}
be fundamental cycles of 
\begin{math}
\pi \mathrm{G}_0
\end{math}
whose endpoints are 
\begin{math}
u_0 = \pi \circ \pi' \left( \left( u_0, \sigma_0 \right) \right)
\end{math}.
By using Proposition 1A.2 in Ref.~\citenum{Hatcher},\begin{math}
\pi_1 \left( \pi \mathrm{G}_0, u_0 \right)
\end{math}
is a free group with basis the classes 
\begin{math}
\left[ C_1 \right], \left[ C_2 \right], \cdots, \left[ C_{N_c} \right]
\end{math}. Consider their lifts and let 
\begin{math}
\left( u_0, \sigma_1 \right), \cdots, \left( u_0, \sigma_{N_c} \right)
\end{math}
be their endpoints. In the case of GRRM, 
\begin{math}
\sigma_i
\end{math}
can be obtained by \verb|*_TS*.log| files by investigating which permutation occurs along each reaction path and multiplying the permutation along each fundamental cycle. By choosing the reference structure appropriately, we can assume 
\begin{math}
\sigma_0
\end{math}
is the identity. 

In this setting, 
\begin{math}
\mathrm{Sym}^{\mathrm{c}}_\Omega \left( v_0 \right)
\end{math}
for 
\begin{math}
v_0 = \pi' \left( u_0, \sigma_0 \right)
\end{math}
is the group generated by 
\begin{math}
\sigma_i \; \left( i \in \left\{ 1, \cdots, N_c \right\} \right)
\end{math}
and the elements of
\begin{math}
\mathfrak{U} \left( r_0 \right)
\end{math}
where 
\begin{math}
v_0 = \left[ r_0 \right]
\end{math}. Since it is obvious that 
\begin{math}
\sigma_i \in \mathrm{Sym}^{\mathrm{c}}_\Omega \left( v_0 \right)
\end{math}
and 
\begin{math}
\mathfrak{U} \left( r_0 \right) \subset \mathrm{Sym}^{\mathrm{c}}_\Omega \left( v_0 \right)
\end{math}
hold, it is enough to show that 
\begin{math}
\sigma_i \; \left( i \in \left\{ 1, \cdots, N_c \right\} \right)
\end{math}
and the elements of 
\begin{math}
\mathfrak{U} \left( r_0 \right)
\end{math}
generate 
\begin{math}
\mathrm{Sym}^{\mathrm{c}}_\Omega \left( v_0 \right)
\end{math}. Take an arbitrary
\begin{math}
\sigma \in \mathrm{Sym}^{\mathrm{c}}_\Omega \left( v_0 \right)
\end{math}. Then, 
\begin{math}
v_0
\end{math}
and 
\begin{math}
\sigma \cdot v_0
\end{math}
are in the same connected component by definition. Therefore, there exists a path
\begin{math}
P
\end{math}
that connects 
\begin{math}
v_0
\end{math}
and 
\begin{math}
\sigma \cdot v_0
\end{math}. Its projection 
\begin{math}
\pi P
\end{math}
is a cycle in 
\begin{math}
\pi \mathrm{G}_0
\end{math}
whose endpoint is 
\begin{math}
u_0 = \pi_{\mathrm{V}} \left( v_0 \right)
\end{math}. The homotopy class of the cycle 
\begin{math}
\left[ \pi P \right]
\end{math}
is a product of 
\begin{math}
\left[ C_1 \right], \cdots, \left[ C_{N_c} \right]
\end{math}. Set 
\begin{equation}
\left[ \pi P \right] = \left[ C_{i_1} \right]^{\epsilon_1} \left[ C_{i_2} \right]^{\epsilon_2} \cdots \left[ C_{i_m} \right]^{\epsilon_m}
\end{equation} 
where 
\begin{math}
i_1, \cdots, i_m \in \left\{ 1, \cdots, N_c \right\}
\end{math}
and 
\begin{math}
\epsilon_1, \cdots, \epsilon_m \in \left\{ -1, 1 \right\}
\end{math}.
Let
\begin{math}
\tilde{P}
\end{math}
be the lifted path in 
\begin{math}
\tilde{\mathrm{G}}_0
\end{math}
of 
\begin{math}
\pi P
\end{math}
starting from 
\begin{math}
\left( u_0, \sigma_0 \right)
\end{math}. Then,
\begin{math} 
\pi' \tilde{P} = P
\end{math}
holds. Since the endpoint of 
\begin{math}
\tilde{P}
\end{math}
is 
\begin{math}
\left( u_0, \sigma_{i_1}^{\epsilon_1} \cdot \sigma_{i_2}^{\epsilon_2} \cdots \sigma_{i_m}^{\epsilon_m} \right)
\end{math},
\begin{math}
\pi' \left( u_0, \sigma_{i_1}^{\epsilon_1} \cdot \sigma_{i_2}^{\epsilon_2} \cdots \sigma_{i_m}^{\epsilon_m} \right) = \sigma \cdot v_0
\end{math}
holds. This implies that 
\begin{math}
\sigma \in \sigma_{i_1}^{\epsilon_1} \cdot \sigma_{i_2}^{\epsilon_2} \cdots \sigma_{i_m}^{\epsilon_m} \mathfrak{U} \left( r_0 \right)
\end{math}
holds. This proves the claim.

\section{GAP program to reproduce the RRM of the isomerization in Fig.2 in the main text} \label{sec:GAP}
This section describes the GAP program to reproduce the isomerization shown in Fig.~2 in the main text. For details on the GAP program, see Ref.~\citenum{GAP4}.

\begin{lstlisting}[language=GAP,caption=Program to compute RRM of bi-tetrahedron isomerization reaction,label=GAPprogram]
sym:=SymmetricGroup(5);
ur:=Subgroup(sym,[(2,3,4),(1,5)(3,4)]);
urt:=Subgroup(sym,[(3,4)(2,5)]);
coseturt:=RightCosets(sym,urt);

edge:=[];
for p in coseturt do
  Append(edge,
    [[CanonicalRightCosetElement(ur,Representative(p)),
       CanonicalRightCosetElement(ur,(Representative(p)^-1*(2,5)(3,4))^-1)]]);
od;
\end{lstlisting}
In this program, we define 
\begin{math}
\mathrm{Sym}_\Omega
\end{math}
in the first line, which is 
\begin{math}
\mathfrak{S}_5
\end{math}
in this case since the five particles are of identical type. In the second and third line, we define the subgroups 
\begin{math}
\mathfrak{U} \left( r_{\mathrm{EQ}} \right)
\end{math}
and 
\begin{math}
\mathfrak{U} \left( r_T \right)
\end{math}
of 
\begin{math}
\mathrm{Sym}_\Omega
\end{math}. In the fourth line, we compute the right coset decomposition of 
\begin{math}
\mathrm{Sym}_\Omega
\end{math}
by 
\begin{math}
\mathfrak{U} \left( r_T \right)
\end{math}. In this computation, we identify each permuntation isomer of the bi-tetrahedron 
\begin{math}
r_{\mathrm{EQ}}
\end{math}
in the shape space to the cosets in Eq.~(28) in the main text. Note that the current version of GAP program does not support the left coset decomposition and thus we compute the right coset decomposition instead and obtain the corresponding left coset decomposition by inverting the representatives of the right cosets. From the 6th line to the 10th line, we compute the edge set of the RRM of the bi-tetrahedron in the shape space. For each right coset \verb|p| of 
\begin{math}
\mathfrak{U} \left( r_T \right)
\end{math}
we take its representative 
\begin{math}
\left( \sigma^\ddag \right)^{-1}
\end{math}
and find the canonical representative of the right coset of 
\begin{math}
\mathfrak{U} \left( r_{\mathrm{EQ}} \right)
\end{math}
containing 
\begin{math}
\left( \sigma^\ddag \right)^{-1}
\end{math}
in the 9th line. In that case, 
\begin{math}
\sigma^\ddag
\end{math}
is a representative of the left coset corresponding to \verb|p| and the inverse of the canonical representative is a representative of the left coset of 
\begin{math}
\mathfrak{U} \left( r_{\mathrm{EQ}} \right)
\end{math}
containing 
\begin{math}
\sigma^\ddag
\end{math}. In the 10th line, we find the canonical representative of the right coset of 
\begin{math}
\mathfrak{U} \left( r_{\mathrm{EQ}} \right)
\end{math}
containing 
\begin{math}
\left( \sigma^\ddag \left( 2, 5 \right) \left( 3, 4 \right) \right)^{-1}
\end{math}. In that case, the inverse of the canonical representative is a representative of the left coset of 
\begin{math}
\mathfrak{U} \left( r_{\mathrm{EQ}} \right)
\end{math}
containing 
\begin{math}
\sigma^\ddag \left( 2, 5 \right) \left( 3, 4 \right)
\end{math}. The resulting list \verb|edge| is the list of pairs of the canonical representatives whose inverses are representatives of the left cosets Eq.~(28).
\section{Implementation detail of the computation of RRM in the shape space from outputs of GRRM program} \label{sec:implementation}
This section presents our implementation for RRMs in Sec.~6 in the main text, which is in Ref~\citenum{reproduce_rrm}. The entire program is governed by bash as shown below. The flowchart of the entire program is in Fig.~\ref{fig:flowchart}.
\begin{figure}[ht]
\begin{tabular}{c}
\begin{tikzpicture}[node distance=2cm]

\node (start) [startstop] {Start};
\node (in1) [io, below of=start] {Input: output files of GRRM, \verb|*_EQ_list.log|, \verb|*_TS_list.log|, and  \verb|*_TS*.log|};
\node (pro1) [process, below of=in1] {Duplicate equilibrium and transition state structures if they are chiral};
\node (pro2) [process, below of=pro1] {Compute $\mathfrak{U} \left( r \right)$ for each reference equilibrium and transition state structure $r$};
\node (pro3) [process, below of=pro2] {For each reaction path corresponding to each transition state, identify the endpoints reference equilibrium structures};
\node (pro4) [process, below of=pro3] {Compute generators of $\textnormal{Sym}_\Omega^c \left( v_0 \right)$};
\node (pro5) [process, below of=pro4] {Compute RRM in shape space by the algorithm in Sec.~3.2};
\node (out1) [io, below of=pro5] {Output: output the RRM in the shape space in the dot format};
\node (stop) [startstop, below of=out1] {Stop};

\draw [arrow] (start) -- (in1);
\draw [arrow] (in1) -- (pro1);
\draw [arrow] (pro1) -- (pro2);
\draw [arrow] (pro2) -- (pro3);
\draw [arrow] (pro3) -- (pro4);
\draw [arrow] (pro4) -- (pro5);
\draw [arrow] (pro5) -- (out1);
\draw [arrow] (out1) -- (stop);

\end{tikzpicture}
\end{tabular}
\caption{Flowchart of the entire program to compute RRM in shape space from outputs of GRRM program}
\label{fig:flowchart}
\end{figure}

The implementation uses output files of GRRM, \verb|*_EQ_list.log|, \verb|*_TS_list.log|, and  \verb|*_TS*.log|. In the 13th line of the program, we extract the list 
\begin{math}
\left\{ \left( r_T^{\left( i \right)}, \left\{ r_R^{\left( i \right)}, r_P^{\left( i \right)} \right\} \right) \right\}_{i \in \tilde{I}}
\end{math}, and add the inversion of the equilibrium state structures and the transition state structures to the list if they are chiral. Thereafter, we compute 
\begin{math}
\mathfrak{U} \left( r \right)
\end{math}
for each reference equilibrium and transition state structure 
\begin{math}
r
\end{math}
using pymatgen symmetry analyzer, a python interface of Spglib: a software library for crystal symmetry search \cite{Togo2018}. Next, for each transition state structure, we identify the endpoints the corresponding reaction path with the reference equilibrium structures along with the necessary permutation to match them using pymatgen.  Thereafter, using the proposed algorithm in Sec.~1, we compute generators of 
\begin{math}
\mathrm{Sym}^{\mathrm{c}}_\Omega \left( v_0 \right)
\end{math}
for a class
\begin{math}
v_0
\end{math}
represented by the equilibrium configuration \verb|EQ0| in \verb|*_EQ_list.log|. In our implementation, a minimum spanning tree of the original RRM graph is identified using NetworkX \cite{networkx}, and fundamental cycles are computed using the minimum spanning tree, and are connected to the vertex 
\begin{math}
v_0
\end{math}
in Sec.~1. The results are inputted into our GAP program \verb|generate_rrm_v9.g| in the line 15--23 of the program and the resulting RRM in the shape space is outputted in the dot file format in the line 25--35 and in the visualized graph  using Graphviz \cite{10.1007/3-540-45848-4_57} in the 37th line. For details of implementation of 
\verb|generate_rrm_v9.g| and \verb|rrm_reconstruction_v12.py|, refer to Ref.~\citenum{reproduce_rrm}.
\begin{lstlisting}[language=bash,caption=Program to extract a relevant information from the output files of GRRM,label=GRRMprogram]
#! /bin/sh

# Specify the following log files of GRRM 
EQLIST="***_EQ_list.log"
TSLIST="***_TS_list.log`" 
# output files including irc information
# (file names are supposed to be "***_TSn.log" where n is a non-negative integer)
TSHEAD="***_TS"

# output file to feed it into GAP
GLIST="data/${NAME}.g"

python rrm_reconstruction_v12.py $EQLIST $TSLIST $TSHEAD $GLIST 

/usr/local/gap-4.11.1/bin/gap.sh -b -q -m 12g generate_rrm_v9.g << EOI
Read("$GLIST");;
vlabel:=true;;
elabel:=true;;

Print("the index of symc in sym : ",IndexNC(sym,symc),"\n");
generate_rrm("$VFILE","$EFILE",symc,ur,urt,ss,org_eq,org_ts,vlabel,elabel);
QUIT;
EOI

# output the RRM in the shape space in the dot format
echo "graph $NAME {" > $OFILE
echo "        outputorder=\"edgesfirst\"" >> $OFILE
echo "        overlap=false" >> $OFILE
echo "        spline=true" >> $OFILE
echo "        frontname=\"Helvetica,Arial,sans-serif\"" >> $OFILE
echo "        node [fontname=\"Helvetica,Arial,sans-serif\"]" >> $OFILE
echo "        edge [fontname=\"Helvetica,Arial,sans-serif\",color=lightgray]" >> $OFILE
cat $VFILE >> $OFILE
cat $EFILE >> $OFILE
echo "}" >> $OFILE

dot -Kfdp -Tpng $OFILE -o $GFILE

echo "DONE"
\end{lstlisting}

\bibliography{dyn_recomb}